\documentclass[aps,prc,twocolumn,showpacs,floatfix,nofootinbib]{revtex4} 
\usepackage{graphicx}
\usepackage{dcolumn}
\usepackage{amsmath}
\usepackage[T1]{fontenc}
\usepackage{txfonts}
\usepackage{booktabs}
\usepackage[pdftex]{hyperref}
\usepackage[applemac]{inputenc}

\usepackage[usenames,dvipsnames]{color} 

\begin{document}

\title{Level densities and $\gamma$-ray strength functions in Sn isotopes}

\author{H.~K.~Toft$^1$\footnote{Email address: h.k.toft@fys.uio.no}, A. C.~Larsen$^1$, U.~Agvaanluvsan$^{2,3}$, A.~B\"urger$^1$, M.~Guttormsen$^1$, G.~E.~Mitchell$^{4,5}$, H. T.~Nyhus$^1$, A.~Schiller$^6$, S.~Siem$^1$, N.~U.~H.~Syed$^1$, 
and A.~Voinov$^6$ }

\affiliation{$^1$Department of Physics, University of Oslo, N-0316 Oslo, Norway}
\affiliation{$^2$Stanford University, Palo Alto, California 94305, USA}
\affiliation{$^3$MonAme Scientific Research Center, Ulaanbaatar, Mongolia}
\affiliation{$^4$Department of Physics, North Carolina State University, Raleigh, NC 27695, USA}
\affiliation{$^5$Triangle Universities Nuclear Laboratory, Durham, NC 27708, USA}
\affiliation{$^6$Department of Physics, Ohio University, Athens, OH 45701, USA}

\date{\today}

\begin{abstract}
The nuclear level densities of $^{118,119}$Sn and the $\gamma$-ray  strength functions of $^{116,118,119}$Sn below the neutron separation energy are extracted with the Oslo method using the ($^3$He, \,$\alpha \gamma$) and ($^3$He,$^3$He$^\prime\gamma$) reactions.  
The level density function of $^{119}$Sn displays step-like structures. 
The microcanonical entropies are deduced from the level densities, and the  single neutron entropy of $^{119}$Sn is determined to be $(1.7 \pm 0.2)\,k_B$. 
Results from a  combinatorial model  support the interpretation  that some of the low-energy steps in the level density function are caused by  neutron pair-breaking. 
An enhancement in all the $\gamma$-ray strength functions of $^{116-119}$Sn, compared to standard models for radiative strength, is observed for the $\gamma$-ray energy region of $\simeq (4 -11)$ MeV. These small resonances all have a centroid energy of 8.0(1) MeV and an integrated strength corresponding to $1.7(9)\%$ of the  classical Thomas-Reiche-Kuhn  sum rule. 
The Sn resonances   may be due to electric dipole neutron skin oscillations or to an enhancement of the giant magnetic dipole resonance.

\end{abstract}  

\maketitle

\section{Introduction}
The level density and the $\gamma$-ray strength function are average quantities describing atomic nuclei. They are important for many aspects of fundamental and applied nuclear physics, including calculations of nuclear properties, like reaction cross sections. Such cross sections are used for calculations in, e.g., reactor physics and nuclear waste management, and of nuclear reaction rates in astrophysics for modeling of nucleosynthesis in stars. 

The nuclear level density of nuclei is defined as the  number of levels per unit of excitation energy.  The entropy and other thermodynamic properties may also be determined  from the level density.
Structures in the level density are expected to be due to shell gaps, breaking of nucleon Cooper pairs, and/or changes in the nuclear shape.

In the majority of previous experiments, the level density is measured either only at low energy by direct counting (conventional spectroscopy), or at higher energy around the neutron/proton separation energies (nuclear resonance measurements). 

The $\gamma$-ray strength function may be defined as the reduced average transition probability as a function of $\gamma$-ray energy. This quantity characterizes average electromagnetic properties of excited nuclei. 
The strength function reveals essential information about the nuclear structure. Electric transitions are mostly influenced by the proton charge distribution,  while magnetic transitions are also affected by the neutron distribution due to the magnetic dipole moment of the neutron. The shape and softness of the nuclear surface are other important factors for the nuclear response to electromagnetic radiation. 
 
The large number of stable  isotopes in Sn makes the element suitable for systematic studies.
This paper presents
the  level densities of $^{118,119}$Sn and the $\gamma$-ray strength functions of  $^{116,118,119}$Sn for energies in the quasi-continuum below the neutron separation energy. The measurements   have been performed at the Oslo Cyclotron Laboratory (OCL). A combinatorial  model is  also  used in order to study, e.g., the origin of the level density steps and the  impact of collective effects.

The $^{118,119}$Sn results are compared with earlier OCL studies on other isotopes.  In Ref.~\cite{Sn_Density}, the level density functions of $^{116,117}$Sn were shown to display steps that are much more distinctive than previously measured for other mass regions. The steps were interpreted as  neutron pair-breaking. 
In Ref.~\cite{Sn_Strength}, a resonance-like structure in the $\gamma$-ray strength function was measured  below the neutron threshold in $^{117}$Sn.

The experimental set-up and the data analysis are briefly described in Sec.~II. The normalized experimental results for level density and entropy  are presented in Sec.~III. Section IV discusses  the nuclear properties extracted from the level density with the combinatorial model. Section V presents the  normalized experimental $\gamma$-ray strength functions. Conclusions are drawn in Sec.~VI.

\section{Experimental set-up and data analysis} 
\label{sec:setup}

The self-supporting $^{119}$Sn target 
was enriched to $93.2\%$ and
had a mass thickness of  1.6~mg/cm$^{2}$. For three days the target was exposed to a 38-MeV $^3$He beam with an average current of $\sim$ 1.5 nA.
The reaction channels studied were $^{119}$Sn($^3$He,$^3$He$^\prime\gamma$)$^{119}$Sn and $^{119}$Sn($^3$He,\,$\alpha\gamma$)$^{118}$Sn.

Particle-$\gamma$ coincidences were recorded with 7 collimated Si particle $\Delta E-E$ telescopes and  26 collimated NaI(Tl) $\gamma$-ray detectors. The $\Delta E$  and $E$ detector thicknesses were about 140~$\mu$m and 1500 $\mu$m, respectively. These detectors were placed at 45$^\circ$ with respect to the beam axis. The NaI detectors 
are distributed on a sphere and constitute the  CACTUS multidetector system \cite{Gut96}.
The total solid-angle coverages out of 4$\pi$ were approximately 1.3$\%$ for the particle detectors and 16$\%$ for the $\gamma$-ray detectors. 

In the data analysis, the measured ejectile's energy is transformed into excitation energy of the residual nucleus using reaction kinematics. 
The $\gamma$-ray spectra for various initial excitation energies are unfolded with the known response functions of CACTUS and the Compton subtraction method  \cite{Gut96}. The Compton subtraction method preserves the fluctuations in the original spectra without introducing further, spurious fluctuations.

The first generation 
$\gamma$-ray spectra are extracted from the unfolded total  $\gamma$-ray spectra, by the subtraction procedure described in Ref.~\cite{Gut87}. The main assumption is that the $\gamma$-decay from any excitation energy bin is independent  of the method of formation -- whether it is directly formed by a nuclear reaction or indirectly  by  $\gamma$-decay  from higher lying states following the initial reaction. 

The first generation $\gamma$-ray spectra are arranged in a 2-dimensional matrix $P(E, E_\gamma)$. 
The entries of $P$ are the probabilities $P(E,E_\gamma)$ that a $\gamma$-ray of energy $E_\gamma$ is emitted from an energy bin of excitation energy $E$. This matrix is used for the simultaneous extraction of the $\gamma$-ray strength function and the level density function. 

The first generation matrix $P$
is factored into the level density function $\rho$ and the radiative transmission coefficient ${\cal T}$  \cite{Sch00a}:
\begin{equation}
\label{probab}
P(E,E_\gamma)\propto {\cal T}(E_\gamma)\, \rho(E-E_\gamma)\,.
\end{equation}
The factorization of $P$ into two components is justified for nuclear reactions leading to a compound state prior to a subsequent $\gamma$-decay \cite{Bohr-Mottelson}. Equation (\ref{probab}) may also be regarded as a variant of Fermi's golden rule: The decay rate is proportional to the density of the final state and the square of the matrix element between the initial and final state. The factorization is performed by an iterative procedure where the independent functions $\rho$ and ${\cal T}$ are adjusted until a global $\chi^2$ minimum with  the experimental $P(E,E_\gamma)$ is reached. 

As shown in Eq.~(\ref{probab}),
the transmission coefficient is assumed to be a function of only $E_\gamma$, in accordance with  the 
 generalized form of the Brink-Axel hypothesis~\cite{Bri55,Axe62}. This hypothesis states that a giant electric dipole resonance, and all other collective excitation modes, may  be built on any excited state and still have the same properties as the one built on the ground state. Hence, the  transmission coefficient is independent of excitation energy. 

Equation (\ref{probab}) determines only the  functional forms of $\rho$ and $\cal T$. The entries of $P$ are invariant under the following transformations \cite{Sch00a}:
\begin{eqnarray}
\tilde{\rho}(E-E_\gamma)&=&A\exp\left[\alpha\left(E-E_\gamma\right)\right]\,\rho(E-E_\gamma)\,,
\label{eq:array1}\\
\tilde{\cal T}(E_\gamma)&=&B\exp\left(\alpha E_\gamma\right) \,{\cal T}(E_\gamma)\,.
\label{eq:array2}
\end{eqnarray}
The  final step of the Oslo method is to determine the normalization parameters. The parameters $A$ and $B$ will define the absolute values of $\rho$ and ${\cal T}$, respectively, while $\alpha$ will define their common slope.

\section{Level densities} 
\label{LD}

\subsection{Normalization and experimental results}
\label{LD-results}
The constants $A$ and $\alpha$ in Eq.~(\ref{eq:array1}), which are needed to normalize the experimental level density $\rho$, are determined using literature values of the  known discrete energy levels at low energy and of  the level  spacing $D$ at the neutron separation energy $S_n$, obtained from neutron resonance experiments. 

The normalization value  $\rho(S_n)$ is calculated 
 either from the $s$-wave level spacing $D_0(S_n)$ or from the $p$-wave level spacing $D_1(S_n)$. 
 The level spacings are taken from Refs.~\cite{Mughabghab,RIPL-3}. 
 To establish an expression for the value of $\rho(S_n)$, it is necessary to assume models for the spin distribution $g(E,I)$ and the spin cutoff parameter $\sigma$.
 We choose the back-shifted Fermi gas (BSFG) model with the original parameterization of von Egidy {\em et al.}  \cite{Egi88}, because this parameterization gives the most appropriate normalization of these nuclei when comparing to other experimental measurements (see also Ref.~\cite{Sn_Density}).
Here,  these functions are kept as the original Gilbert and Cameron  expressions \cite{Gil65}, but with a redefined  parameterization of the nucleus' intrinsic excitation energy $U$ and the level density parameter $a$. The spin distribution is  expressed as  \cite{Egi88}:
\begin{equation}
\label{eqn:GC-spin}
g(E,I)\simeq \frac {2I+1}{2\sigma^2}\exp{\left[ -\left(I+1/2\right)^2/2\sigma^2\right]}\,,
\end{equation}
where $I$ is the spin, and where the spin cutoff parameter $\sigma(E)$ is given by:
 \begin{equation}
 \label{eqn:cutoff}
 \sigma^{2}~=~0.0888 \, A^{2/3} aT\,,
 \end{equation}
where $A$ is the mass number of the isotope, and $T$ is the nuclear temperature given by $T = \sqrt{U/a}$.  Here, the level density parameter is defined as
$a~=~0.21 \, A^{0.87}~{\rm MeV^{-1}}$, while
the shifted excitation energy $U$ is defined as $U = E - E_{\rm pair} - C_1$. The  back-shift parameter is defined as $C_1 = -6.6 \, A^{-0.32}$ MeV.
The pairing energy $E_{\rm pair}$ is calculated from the proton and neutron pair-gap parameters: $E_{\rm pair} = \Delta_{\rm p} +\Delta_{\rm n}$. 
The pair-gap parameters are evaluated from the even-odd mass differences found in Ref.~\cite{Wapstra} according to the method of Ref.~\cite{Dob01}. 

 Assuming this spin distribution and equal numbers of levels with positive and negative parity, 
  the level density at $S_n$  may be expressed as, for $s$-wave neutron resonances \cite{Syed09, Sch00a}:
 \begin{eqnarray}
 \label{eq:D0}
\rho_0(S_n) &=& \frac{2{\sigma}^2}{D_0}\cdot \left\{\left(I_t + 1\right)\exp\left[ \frac{-\left(I_t +1\right)^2}{2\sigma^2}\right]  \right. \nonumber \\
  &+& \left. I_t  \exp\left[ \frac{-{I_t}^2}{2\sigma^2}\right] \right\}^{-1} \,,
 \end{eqnarray}
 and for $p$-wave resonances \cite{Syed09}:
 \begin{eqnarray}
 \label{eq:D1}
 \rho_1(S_n) &=& \frac{2\sigma^2}{D_1} \cdot \left\{ \left(I_t-1\right)\exp \left[ \frac{-\left(I_t -1\right)^2}{2\sigma^2}\right]  +  I_t \exp\left[ \frac{-{I_t}^2}{2\sigma^2}\right]  \right. \nonumber \\
 & + & \left. \left(I_t +1\right)\exp\left[\frac{ -\left(I_t +1\right)^2}{2\sigma^2}\right]  \right. \nonumber \\
  & + & \left. \left(I_t +2\right) \exp\left[ \frac{-\left(I_t +2\right)^2}{2\sigma^2}\right]   \right\} ^{-1}  \,,
 \end{eqnarray}
 where  the spin cutoff parameter is evaluated at $S_n$, and where $I_t$ is the spin of the target. 

A higher $\rho(S_n)$ is obtained from the level spacing of $D_0$ than of $D_1$, according to calculations on both isotopes. As the highest value of the level density is presumed to be the best estimate,  $D_0$ is chosen in the following.
The input parameters and the resulting values of the normalization data $\rho(S_n)$ are given in Tab.~\ref{tab:parametre}. 

\begin{table*}[!htb] 
\caption{Input parameters and the resulting values for the calculation of the normalization value $\rho(S_n)$, and the input parameters for the BSFG interpolation and the required values of the scaling parameter $\eta$. }
\begin{tabular}{|l|cccccccc|c|c|}
\hline
\hline
Nucleus     & $S_n$  &  $D_0(S_n)$ & $a$    & $C_1$   & $\Delta_{\rm n}$ & $\Delta_{\rm p}$    & $\sigma(S_n)$  & $\rho$($S_n$) & $\eta$ \\ 
          & (MeV)  &  (eV) &  (MeV$^{-1}$)  & (MeV)   & (MeV) & (MeV)   & &  $(10^{4}$ MeV$^{-1})$  & \\
\hline
$^{119}$Sn  & 6.485 & 700(150) & 13.43 & $ -1.43 $ & 0 & 1.02 &  4.55   & 6.05(175)  & 0.44   \\
$^{118}$Sn  &   9.326  & 61(7)   & 13.33   & $-1.43$ &  1.19 & 1.24 &  4.74   & 38.4(86)   & 0.59 \\
\hline
\hline
\end{tabular}
\label{tab:parametre}
\end{table*}

The experimental data for the level densities are not obtained up to the excitation energy of $S_n$.
There is a gap, and the level density in the  gap and below is estimated according to the  level density prediction of the BSFG model with the parameterization of von Egidy  {\em et al.}  \cite{Egi88}. 
This is a consistency choice in order to keep 
the spin distribution and the spin cutoff parameter  the same as the ones  used during the calculation of $\rho(S_n)$ based on the neutron resonance data. 
The BSFG level density, for
all spins and as a function of excitation energy, is given by
\begin{equation}
\label{eqn:rho}
\rho(E)_{\rm BSFG} =\frac{\mathrm{exp}\left(2\sqrt{aU}\right)}{12\sqrt{2}\,a^{1/4}\,U^{5/4}\,\sigma}\,.
\end{equation}
A scaling parameter $\eta$ is applied to the BSFG formula,
\begin{equation}
\rho(E)_{\rm BSFG}\rightarrow\eta\,\rho(E)_{\rm BSFG}\,,
\end{equation}
in order to make its absolute value at $S_n$ agree with the normalization value $\rho(S_n)$. We then get a level density interpolation that overlaps with the measurements, and to which the measurements are normalized.
The values of $\eta$  are shown in Tab.~\ref{tab:parametre}.  

\begin{figure}[!ht]
\includegraphics[width=9.5cm]{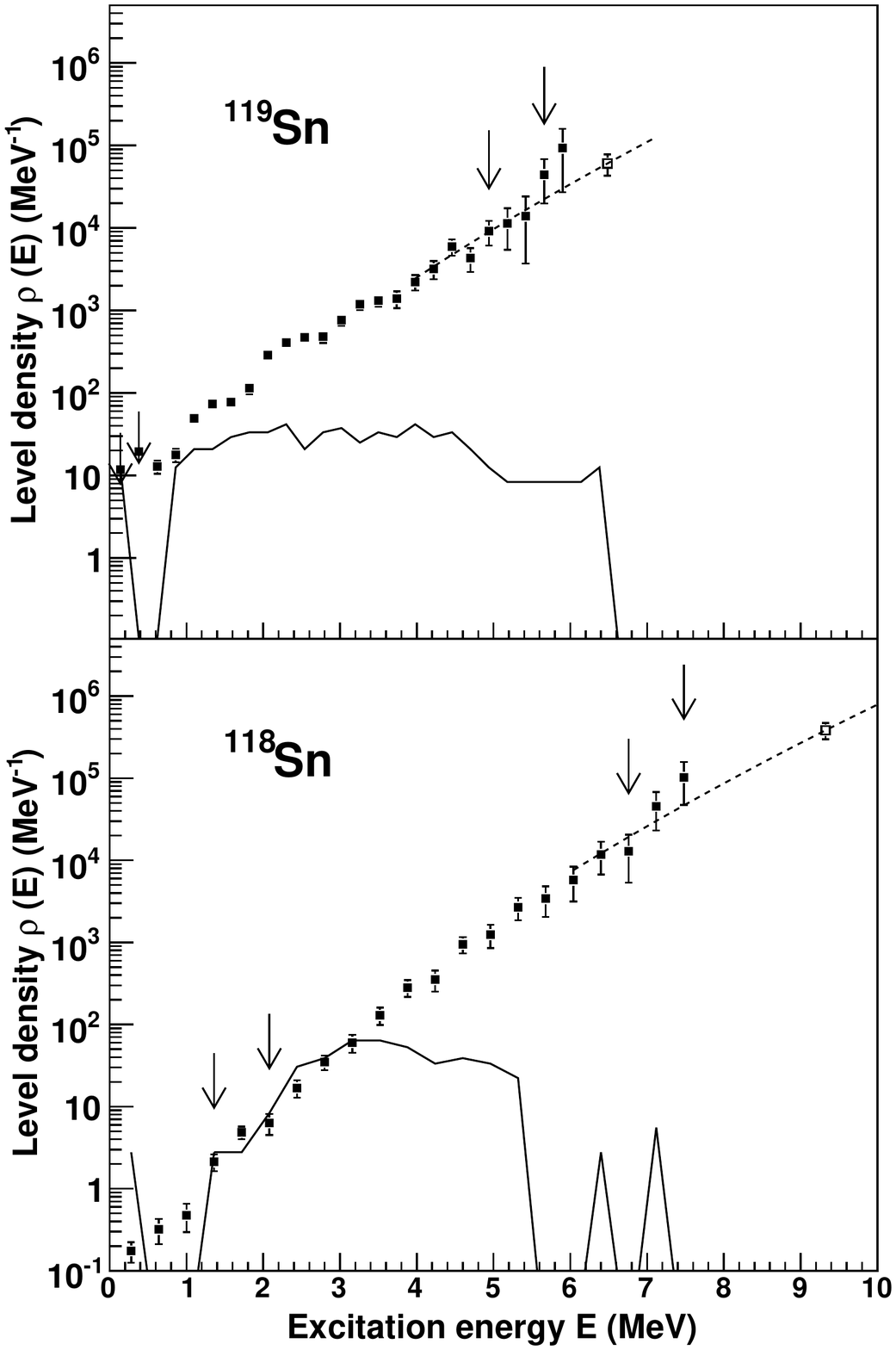}
\caption{Normalized level densities of $^{119}$Sn (upper panel) and $^{118}$Sn (lower panel) as a function of excitation energy. Our experimental data are marked with filled squares. The dashed lines are the BSFG predictions that are used for interpolation, scaled to coincide with  $\rho(S_n)$ (open squares), which are calculated from neutron resonance data. The solid lines represent the discrete level densities  obtained from  counting  the known levels. The arrows indicate the regions used to normalize the absolute values and the slope. The energy bins are 360 and 240 keV/ch  for $^{118,119}$Sn, respectively.}
\label{rhonorm}
\end{figure}

Figure \ref{rhonorm} shows the normalized level densities in $^{118,119}$Sn. The arrows indicate the regions used for normalization. As expected, the level densities of $^{119}$Sn and $^{118}$Sn are very similar to those of $^{117}$Sn and $^{116}$Sn \cite{Sn_Density},  respectively.
 
The figure also shows that the known discrete levels \cite{ToI} seem to be complete up to $\sim$ 2  MeV in $^{119}$Sn and up to $\sim 3$~MeV  in  $^{118}$Sn. 
Hence, our experiment has filled a region of unknown level density from the discrete region and  to  the gap, approximately at $\left( S_n-1\right)$ MeV.
Unlike $^{119}$Sn, the ground-state of the even-even nucleus $^{118}$Sn has no  unpaired neutron,  and accordingly it has fewer available states than $^{119}$Sn. Therefore, measuring  all levels to higher excitation energies by conventional methods is easier in $^{118}$Sn.

\begin{figure}[!ht]
\includegraphics[width= 9.5 cm]{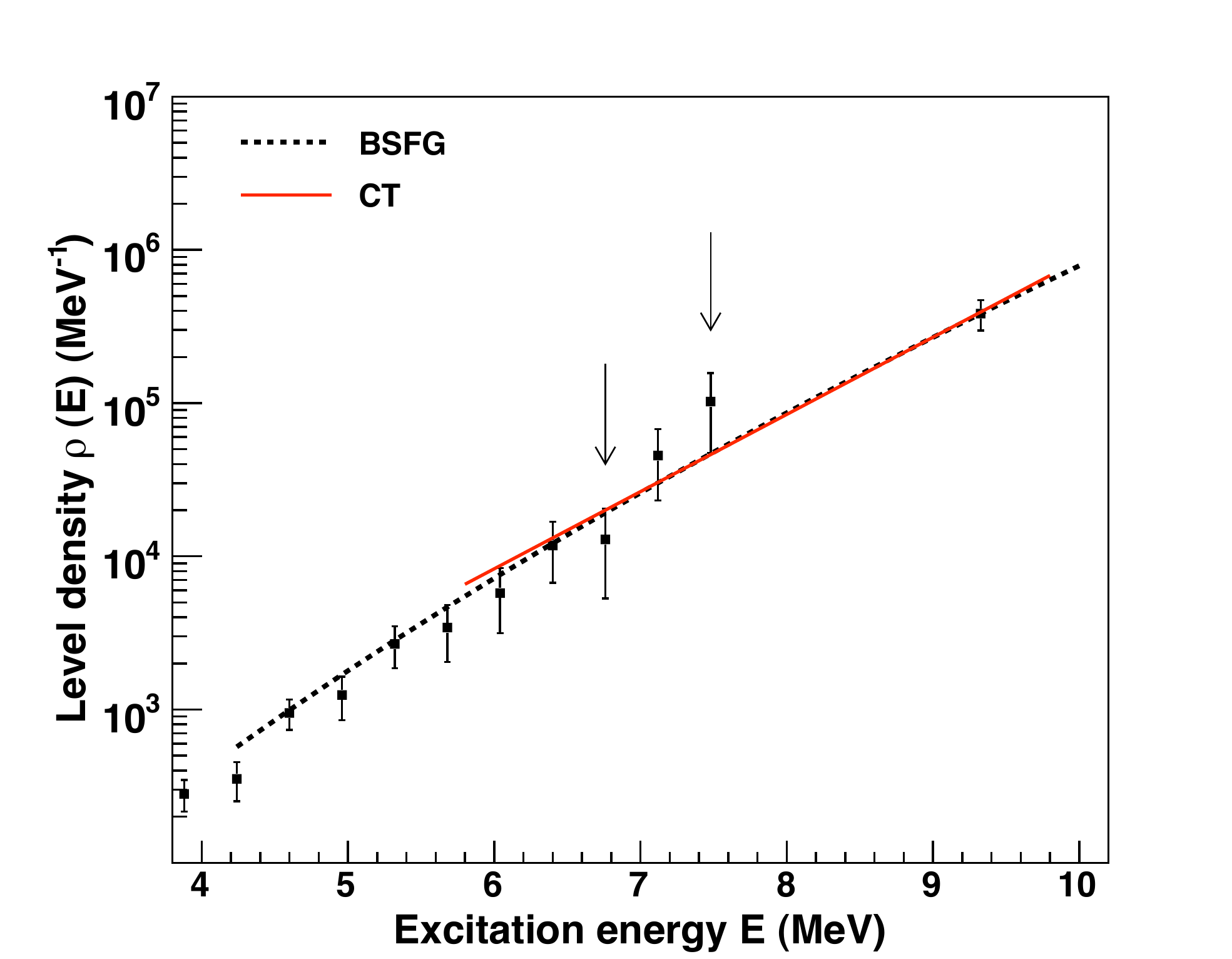}
\caption{(Color online.) Comparison of the BSFG model (dashed line) and the CT model (solid line) as interpolation means  for the level density of $^{118}$Sn. The arrows indicate the region of normalization. The parameters in the CT model ($T= 0.86$ MeV and $E_0 = -1.7$~MeV) have been found from a least $\chi^2$ fit to the data points in this region and from matching $\rho(S_n)$, and the parameterization is not intended to be appropriate elsewhere. }
\label{rhoCTmodel}
\end{figure}

An alternative interpolation method to describe  the gap between our measured data and 
 the neutron resonance data based
$\rho(S_n)$  is the constant temperature (CT) model  \cite{Gil65}. 
This approximation gives 
\begin{equation}
\rho(E)=\frac{1}{T}\exp\left[\left(E-E_0\right)/T\right]\,,
\end{equation}
where the "temperature" $T$ and the energy shift $E_0$ are treated as free parameters.
Figure \ref{rhoCTmodel} shows a comparison of the CT model and the BSFG model as  interpolation methods for $^{118}$Sn. The small difference in the region of interpolation 
is negligible for the normalization procedure.

\subsection{Step-like structures}
\label{subsec-steps}
In Fig.~\ref{rhonorm}, the level density of $^{119}$Sn shows a step-like structure superimposed on the general level density, which is smoothly increasing  as a function of excitation energy. A step is characterized by an abrupt increase of level density within a small energy interval. The phenomenon of steps was also seen in $^{116,117}$Sn \cite{Sn_Density}. 

Distinctive steps 
in $^{119}$Sn are seen below $\sim 4$ MeV. They are, together with the steps of $^{116,117}$Sn  \cite{Sn_Density}, the most distinctive steps measured so far at OCL. This  may be explained by Sn having a magic number of protons, $Z=50$. 
As long as the excitation energy is less than the energy of the proton shell gap, only neutron pairs are broken. The steps are distinctive since  no proton pair-breaking smears the level density function. 

The steps are less pronounced for $^{118}$Sn than for $^{119}$Sn. This is in contradiction to what is expected, as $^{118}$Sn is an even-even nucleus without the unpaired neutron reducing the clearness of the steps in $^{119}$Sn. The explanation probably lies in poorer statistics for the $(^{3}{\rm He},\alpha)$ reaction channel than for $(^{3}{\rm He}, ^{3}{\rm He}^\prime)$. To reduce the error bars, a larger energy bin is chosen  for  $^{118}$Sn, leading to  smearing the data and less clear structures.

Two steps in  $^{119}$Sn are particularly distinctive: one at $\sim~1.0$~MeV and another at $\sim~2.0$~MeV, leading to bumps in the region around $\sim (1.2 - 1.4)$~MeV and around $\sim (2.2 - 2.6)$~MeV, respectively.
The steps in $^{119}$Sn are  found at approximately the same locations as  in $^{117}$Sn \cite{Sn_Density}. 

Also for $^{116}$Sn, two steps were clearly seen for low excitation energy \cite{Sn_Density}. The first of these is probably connected to the isotope's first excited state, at 1.29 MeV  \cite{ToI}.
A similar step in $^{118}$Sn would probably also had been found connected to the first excited state, at 1.23 MeV   \cite{ToI}, if the measured data had had better statistics.

Microscopic calculations based on the seniority model indicate that step structures in  level density functions may be explained by the consecutive breaking of nucleon Cooper pairs~\cite{Fe+Mo_lev}. The steps for $^{119}$Sn in Fig.~\ref{rhonorm}  are probable candidates for the neutron pair-breaking process. 
The neutron pair-breaking energy of $^{119}$Sn is estimated\footnote{The values of the 
neutron pair-breaking $2\Delta_{\rm n}$ and the proton pair-breaking $2\Delta_{\rm p}$
for $^{118,119}$Sn are  estimated from the $\Delta_{\rm n/p}$ values  in Tab.~\ref{tab:parametre}, except for $\Delta_{\rm n}$ of  $^{119}$Sn.  We estimate the energy for breaking a neutron pair in  $^{119}$Sn  as
 the mean value of $2 \Delta_{\rm n}$  of the neighbouring even-even nuclei, redefining  its value to be  $2\Delta_{\rm n} = 2.5$ MeV.} 
to be $2\Delta_{\rm n}=2.5$~MeV, which supports neutron pair-breaking as the origin of the pronounced bump around $\sim (2.2 - 2.6)$ MeV. 

However, if the applied values of the neutron pair-gap parameters are accurate, the pronounced step at $\sim 1.0$~MeV  in $^{119}$Sn and other steps below this energy are  probably not due to pure neutron pair-breaking. They might be due to more complex structures,  involving collective effects such as vibrations and/or rotations.
In Sec.~\ref{sub-pairs}, the pair-breaking in our isotopes will be investigated further.

\subsection{Entropy} 

In many fields of natural science, the entropy is used to reveal  the degree of order/disorder of a system. In nuclear physics, the entropy may describe the number of ways the nucleus can arrange for a certain excitation energy. Various thermodynamic quantities may be deduced from the entropy, e.g., temperature, heat capacity and chemical potential. The study of nuclear entropy also exhibits the amount of entropy gained from the breaking of Cooper pairs.
We would like to study  the entropy difference between  odd-$A$ and even-even Sn isotopes. 

The microcanonical entropy is defined as
\begin{equation}
S_s(E)=k_{\rm B}\ln \Omega_s(E)\,,
\end{equation}
where $k_{\rm B}$ is the Boltzmann constant, which is set to unity to make the entropy  dimensionless, and where $\Omega_s(E)$ is the state density (multiplicity of accessible states).  The state density is proportional to the experimental level density $\rho(E)$ by
\begin{equation}
\label{eq:omega-s}
\Omega_s(E) \propto \rho(E) \cdot \left[ 2\left<I(E)\right> +1 \right]\,,
\end{equation}
where $\left<I(E)\right>$ is the average spin within an energy bin of excitation energy $E$. The factor $2\left<I(E)\right> +1$ is the spin  degeneracy of magnetic substates. 

The spin distribution is not well-known, so we assume  the spin degeneracy factor  to be constant and   omit it. 
Omitting this factor is firstly grounded by
 the  spin being averaged over each energy bin, leading to only the absolute value of the state density at high excitation energies being altered, and not the structure.
 Secondly, the average spin $\left< I(E)\right>$ is expected to be a slowly varying function of energy (see Sec.~\ref{sec:combi}).
Hence, a  "pseudo" entropy $S_l$ may be defined based only on the level density $\rho(E)$:
\begin{equation}
S_l\left(E\right)=k_{\rm B}\ln \left( \frac {\rho(E)}{\rho_0}\right) \,.
\label{eq:pseudoentropi}
\end{equation}
The constant $\rho_0$ is chosen so that $S_l = 0 $ in the ground state of the even-even nucleus $^{118}$Sn. This is  satisfied for $\rho_0 = 0.135$~MeV$^{-1}$. The same value of $\rho_0$ is used  for $^{119}$Sn.

\begin{figure}[tb]
\includegraphics[width=9.5cm]{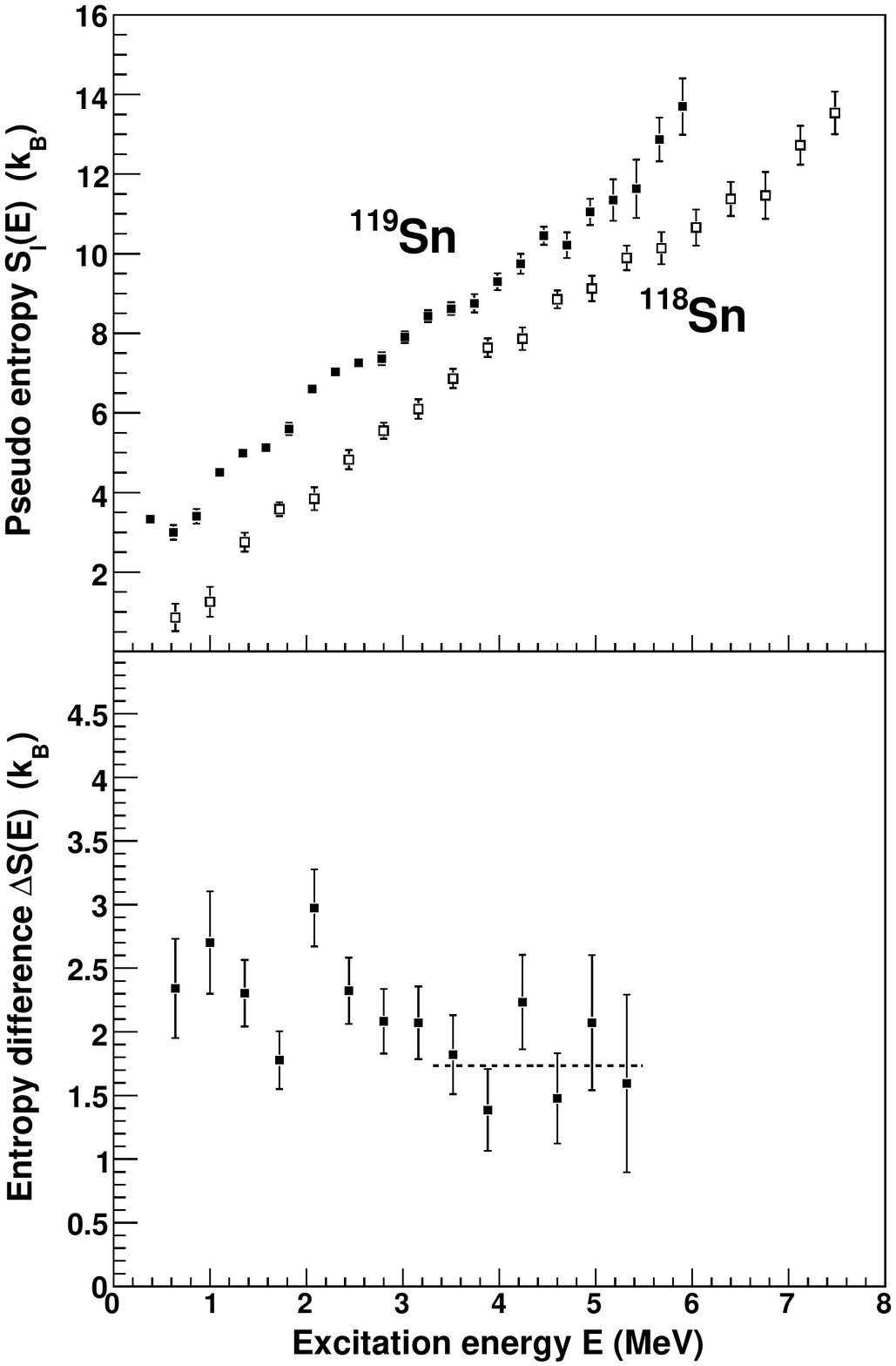}
\caption{Upper panel: Experimental pseudo entropies $S_l$ of $^{119}$Sn (filled squares) and $^{118}$Sn (open squares) as a function of excitation energy. Lower panel: The respective  experimental entropy difference, $\Delta S = {S_l}^{119}-{S_l}^{118}$, as a function of excitation energy. 
An average value of $\overline{\Delta S}(E) = \left(1.7 \pm 0.2 \right) k_B$ is obtained from a $\chi^2$-fit (dashed line) to the experimental data above $\sim 3$ MeV.}
\label{entropi}
\end{figure}

Figure \ref{entropi} shows the experimental results for the pseudo entropies of  $^{118,119}$Sn. These pseudo entropy functions  are very similar to those  of $^{116,117}$Sn \cite{Sn_Density}, which is as expected from the general similarity of the level density functions of these isotopes.

We define the entropy difference as
\begin{equation}
\Delta S(E) = {S_l}^{A}- {S_l}^{A-1}\,,
\end{equation}
where the superscript denotes the mass number of the isotope. 
Assuming that entropy is an extensive quantity, the entropy difference will be equal to the entropy of the valence neutron, i.e.~the experimental single neutron entropy of $^{A}$Sn. 

For midshell nuclei in the rare-earth region, a semi-empirical study \cite{gutt4} has shown that the average single nucleon entropy  is $\overline{\Delta S} \simeq 1.7 ~k_B$. This is true for a wide range of excitation energies, e.g., both for 1 and 7 MeV. Hence for these nuclei, the entropy simply scales with the number of nucleons not coupled in Cooper pairs, and the entropy difference is merely  a simple shift with origin from the pairing energy.

Figure \ref{entropi} also shows  the entropy difference $\Delta S$ of $^{118,119}$Sn, which are  midshell in the neutrons only.
Above $\sim 3$ MeV,  the entropy difference may seem to approach a constant value. In the energy region where the entropy difference might be constant (shown as the dashed line in Fig.~\ref{entropi}), we have calculated its mean value as $\overline{\Delta S}= \left(1.7 \pm 0.2 \right) k_B$. Within the uncertainty, this 
limit
is in good agreement with 
 the general conclusion of the above mentioned semi-empirical study \cite{gutt4}, and with
  the findings for   $^{116,117}$Sn \cite{Sn_Density}.
 For lower excitation energy,  however, Fig.~\ref{entropi} shows that the entropy difference of $^{118,119}$Sn is not a constant, unlike the rare-earth midshell nuclei.  Hence, the  $^{118,119}$Sn isotopes have an entropy difference that is more complicated than  a simple excitation energy shift of the level density functions.

\section{Nuclear properties extracted with a combinatorial BCS model}
\label{sec:combi}

A simple microscopic model \cite{CecSc, SyedTi, MagneProceedings} has been developed for further investigation of the underlying nuclear structure resulting in the measured level density functions.
The model distributes Bardeen-Cooper-Schrieffer (BCS) quasi-particles  on single-particle orbitals to make all possible proton and neutron configurations for a given excitation energy $E$. On each  configuration, collective energy terms from rotation and vibration are schematically added. 
Even though this is a very simple representation of the physical phenomena, this combinatorial BCS model reproduces rather well the experimental level densities. As a consequence, the model  is therefore assumed to be able to predict also other
nuclear properties of the system. 
We are first and foremost interested in investigating the level density  steps, and in investigating the assumption of parity symmetry, used in the normalization processes of the Oslo method.

\subsection{The model}
\begin{figure}[tb]
\includegraphics[width=9.5cm]{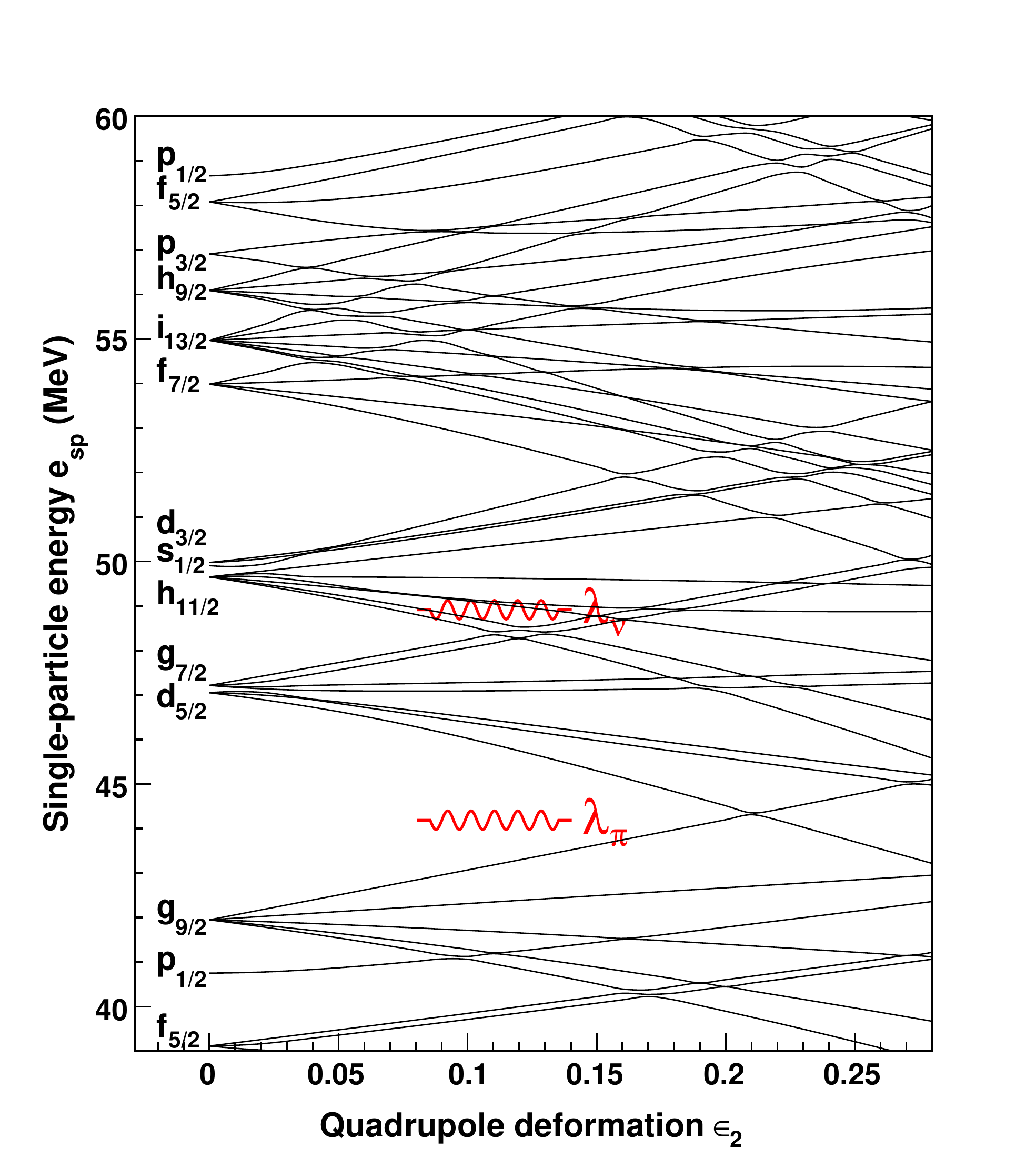}
\caption{(Color online) The Nilsson level scheme, showing single-particle energies as functions of quadrupole deformation $\epsilon_2$, for
$^{118}$Sn, which has $\epsilon_2=0.111$. The  Nilsson parameters are set to $\kappa=0.070$ and $\mu=0.48$. The Fermi levels are illustrated  as curled lines ($\lambda_\pi$ for protons, and $\lambda_\nu$ for neutrons). }
\label{Nilsson}
\end{figure}

The single-particle energies $e_{\rm sp}$ are calculated from the Nilsson model for a nucleus with an axially  deformed core of  quadrupole deformation parameter $\epsilon_2$. 
The values of the deformation parameters are  
 $\epsilon_2 = 0.111$ and $\epsilon_2 = 0.109$ \cite{RIPL-2} for $^{118,119}$Sn, respectively.
Also needed  for the calculation of the Nilsson energy scheme are the Nilsson parameters $\kappa$ and $\mu$ and the oscillator quantum energy  between the main  oscillator shells: $\hbar\omega_0=41A^{-1/3}$.  The adopted values are 
$\kappa=0.070$ and $\mu=0.48$ for both neutrons and protons and for both nuclei, in agreement with the suggestion of Ref.~\cite{KappaMy}. All input parameters are listed in Tab.~\ref{tab:Combi-parametre}. The resulting Nilsson scheme for $^{118}$Sn is shown in Fig.~\ref{Nilsson}. 

\begin{table*}[!htb] 
\caption{Input parameters used in the combinatorial BCS model, and the resulting values for the Fermi levels $\lambda_\nu$ (neutrons) and $\lambda_\pi$ (protons).}
\begin{tabular}{|l|ccccccccc|cc|}
\hline
\hline
Nucleus   & $\epsilon_2$  & $\kappa$  & $\mu$  & $\hbar\omega_0$ & $A_\mathrm{gs}$  & $A_\mathrm{rigid}$ &  $\hbar\omega_{\mathrm{vib,\,e-e}}$ & $\lambda_\nu$  & $\lambda_\pi$    \\ 
       &  & & & (MeV)   & (MeV) &  (MeV) &  (MeV) & (MeV) &  (MeV)  \\
\hline
$^{119}$Sn  &   0.109      & 0.070   &  0.48   &  8.34  &  0.200     & 0.0122   & 1.23; 2.32 &  48.9 & 44.1  \\
$^{118}$Sn   & 0.111     & 0.070     &  0.48   & 8.36  &  0.205      &  0.0124 & 1.23; 2.32 & 48.9 & 44.2   \\
\hline
\hline
\end{tabular}
\label{tab:Combi-parametre}
\end{table*}

The parameter $\lambda$ represents the quasi-particle Fermi level. It is iteratively determined  by reproducing  the right numbers of neutrons and protons in the system. The resulting Fermi levels for our nuclei are listed in Tab.~\ref{tab:Combi-parametre} and  illustrated for $^{118}$Sn in  Fig.~\ref{Nilsson}. 

The microscopic model uses the  concept of BCS quasi-particles \cite{BCS}. Here, the single quasi-particle energies $e_{\rm qp}$ are defined by the transformation of:
\begin{equation}
\label{eqn:BCS}
e_{\rm qp} = \sqrt{\left(e_{\rm sp}-\lambda\right)^2+\Delta^2}\,.
\end{equation}
The pair-gap parameter $\Delta$ is treated as a constant, as before, and with the same values. 

The proton and neutron quasi-particle orbitals are characterized  by their spin projections on the symmetry axes $\Omega_\pi$ and $\Omega_\nu$, respectively. The energy due to quasi-particle excitations is given by the sum of the proton and neutron energies and of a residual interaction $V$: 
\begin{equation}
E_{\rm qp}(\Omega_\pi,\Omega_\nu)=\sum_{\{\Omega^\prime_\pi,\, \Omega^\prime_\nu\}} e_{\rm qp}(\Omega_\pi^\prime) + e_{\rm qp}(\Omega_\nu^\prime) + V(\Omega_\pi^\prime,\Omega^\prime_\nu) \,.
\end{equation}
In the model, quasi-particles having $\Omega$'s of different sign will have  the same energy, i.e.~one has a level degeneracy. Since no such degeneracy  is expected, a  Gaussian random distribution $V$ is  introduced to  compensate for a residual interaction apparently not taken into account by the Hamiltonian of the model.
The maximum allowed number of broken Cooper pairs in our system is 3, giving a total of 7 quasi-particles for the even-odd nucleus $^{119}$Sn. Technically, all configurations are  found from systematic combinations. 

On each configuration, both a vibrational band and rotations are built. The energy of each level is found by adding  the energy of the configuration and the vibrational and rotational terms: 
\begin{equation}
E = E_{\rm qp}(\Omega_\pi,\Omega_\nu)+\hbar\omega_{\rm vib}\nu+A_{\rm rot}R\left(R+1\right) \,.
\label{eqn:energisum}
\end{equation}
The vibrational  term is described by the oscillator quantum energy $\hbar \omega_{\rm vib}$ and the phonon quantum number $\nu = 0, 1, 2, ...$
The values of $\hbar \omega_{\rm vib}$ are found from the $2^+$ and $3^-$ vibrational states of the even-even nucleus and are shown in Tab.~\ref{tab:Combi-parametre}.
The last term of Eq.~(\ref{eqn:energisum}) represents the  rotational energy. The quantity $A_{\rm rot}=\hbar^2/2{\cal J}$ is the rotational parameter with ${\cal J}$ being the moment of inertia, and $R$ is the rotational quantum number. The rotational quantum number has the values of $R=0, 1, 2, 3 ...$ for the even-odd nucleus of $^{119}$Sn, and $R=0, 2, 4 ...$ for the even-even nucleus of $^{118}$Sn.

For low excitation energy, the value of the rotational parameter $A_{\rm rot}$ is determined around the ground state $A_{\rm gs}$. At high  energy,
the rotation parameter is found from  a rigid, elliptical body, which is \cite{Krane}:
\begin{equation}
A_{\rm rigid}=\frac{5\hbar^2}{4MR^2_A \left(1+0.31\epsilon_2\right)}\,.
\end{equation}
Here, $M$ is the mass  and $R_A$ the radius of the nucleus.  
For nuclei  in the medium mass region, $A \sim 50 - 70$, the rotational parameter $A_{\rm rigid}$ is obtained  at the neutron separation energy, according to a theoretical prediction  \cite{Alhassid}. We assume that 
$A_{\rm rigid}$ is obtained  at the neutron separation energy also for our nuclei. The  applied values of  $A_{\rm gs}$ and $A_{\rm rigid}$  are listed in Tab.~\ref{tab:Combi-parametre}. The function $A_{\rm rot}$ as a function of energy is estimated from a linear interpolation between these.

\subsection{Level density}
\begin{figure}[tb]
\includegraphics[width=9.5cm]{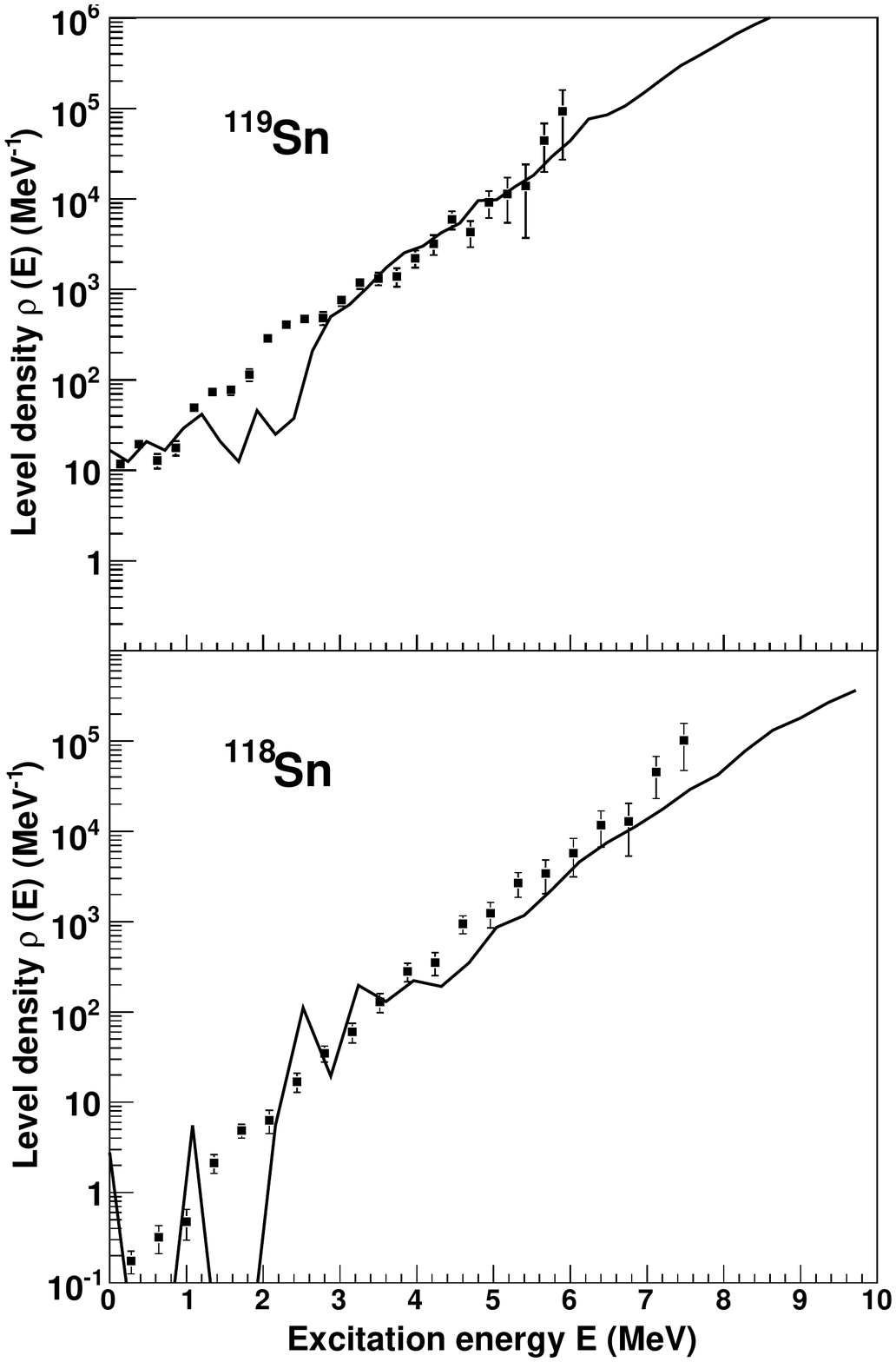}
\caption{Level densities of $^{119}$Sn (upper panel) and $^{118}$Sn (lower panel) as a function of excitation energy. The solid lines are the theoretical predictions of the combinatorial BCS model. The squares are our experimental data.} 
\label{fig:combi-density}
\end{figure}

In Fig.~\ref{fig:combi-density}, the level density functions calculated by our model are compared with the experimental ones. We see that the model gives a very good representation of the level densities in the statistical area above 3 MeV for both isotopes.  
Not taking into account all collective bands known from literature, the model is not intended to reproduce the discrete level structure below the pair-breaking energy.  The model also succeeds in reproducing the bump around $\sim (2.2 - 2.6)$ MeV for both isotopes, even though the onset of this bump in $^{119}$Sn appears to be slightly delayed. 
Above $\sim 2.6$ MeV, the step brings the level density to the same order of magnitude as the experimental values.

According to Eq.~(\ref{eqn:rho}) and the relation between the intrinsic excitation energy $U$ and the pair-gap parameters $\Delta_{\rm p}$ and $\Delta_{\rm n}$ the log-scale slope  of the level density function is dependent on the pair-gap parameters.  Figure \ref{fig:combi-density} shows that the model reproduces well the slopes of the level densities for both isotopes. This supports the applied values of the pair-gap parameters.

\subsection{Pair-breaking}
\label{sub-pairs}

The pair breaking process produces a strong increase in the level density. Typically, a single nucleon entropy of $(1.6 - 1.7) \, k_B $ represents a factor of $\sim 5$ more levels due to the valence neutron. Thus, the breaking of a Cooper pair represents about 25 more levels. Pair-breaking is the most important mechanism for creating entropy in nuclei as function of excitation energy.

\begin{figure}[h!tb]
\includegraphics[width=9.5cm]{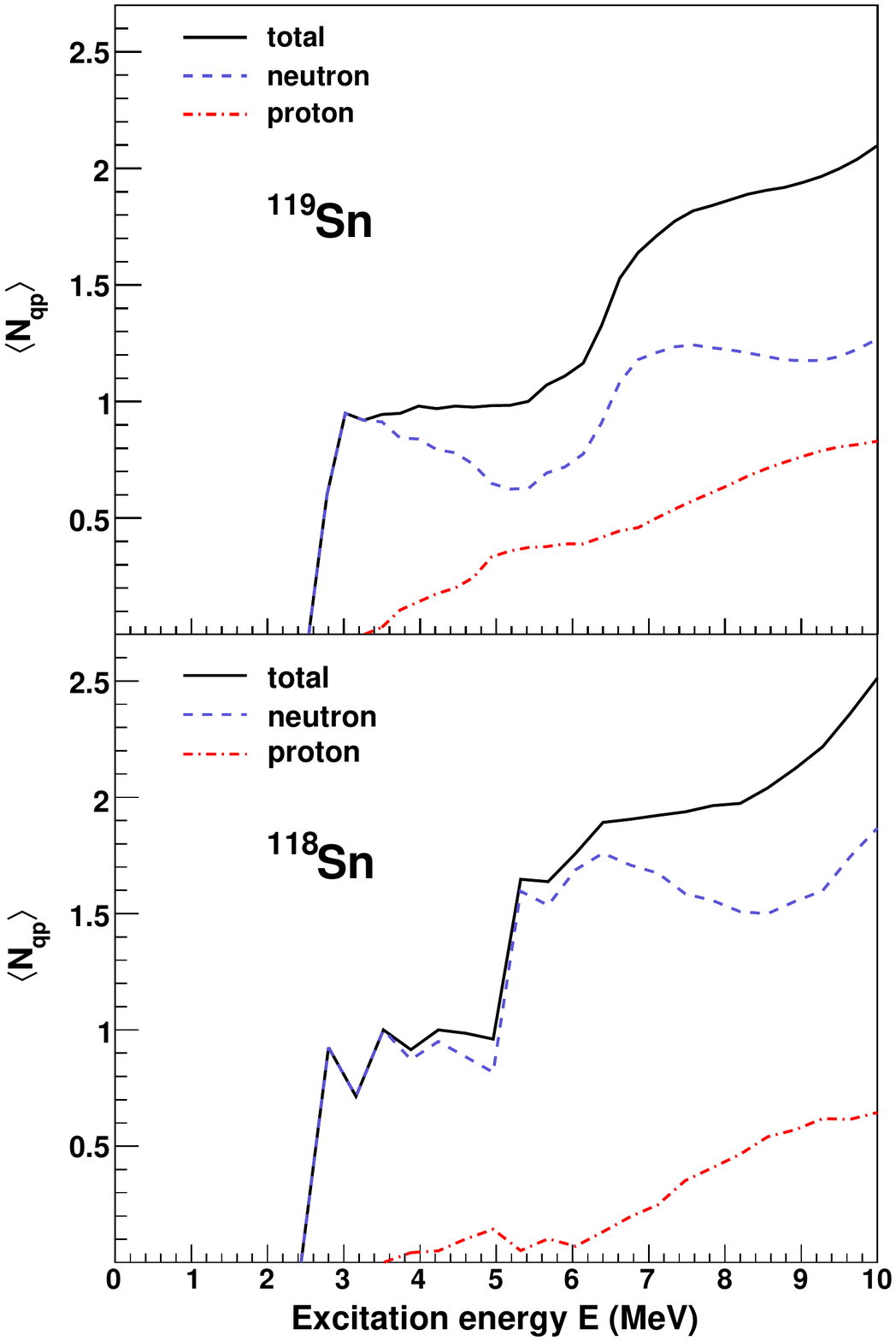}
\caption{(Color online) The average number of broken quasi-particle pairs $\left<N_{qp}\right>$ (solid line) as a function of excitation enegy for  $^{119}$Sn (upper panel) and $^{118}$Sn (lower panel), according to the combinatorial BCS model. Also shown is how this quantity breaks down into neutron pairs (dashed line) and proton pairs (dashed-dotted  line).}
\label{fig:pairs}
\end{figure}

The average number of broken Cooper pairs  per energy bin, $\left<N_{qp}\right>$, is calculated as a function of excitation energy by the model, using  the adopted pair-gap parameters as input values. All configurations obtained for each energy bin are traced, and their respective numbers of broken pairs are counted. The average number of broken pairs is also calculated separately for proton and neutron pairs. 
The result for  $^{118,119}$Sn is shown in Fig.~\ref{fig:pairs}. 

Figure \ref{fig:pairs} shows that the first pair-breaking for both $^{118,119}$Sn is
at an excitation energy around $(2.2 - 2.6)$ MeV. That energy corresponds to the pair-breaking energy plus the extra energy needed to form the new configuration. The figure also shows that, according to the model, the  pair-breakings here are only due to neutrons.
The step in the average number of broken pairs is abrupt, and this number increases from 0 to almost 1. 
This means that there is a very high probability for  the nucleus to undergo a neutron pair-breaking at this energy.
Provided that our values of the pair-gap parameters are reasonable, so that this intense step in the number of broken neutron pairs in the model corresponds
to the distinctive step in level density at $\sim 2.0$ MeV in $^{119}$Sn (see Sec.~\ref{LD}), that  step in level density is probably purely due to  neutron pair-breaking.

The increases in the average number of broken pairs are abrupt also for certain other excitation energies, namely around $(5 - 6)$ MeV and $(8 - 9)$ MeV, as  shown in Fig.~\ref{fig:pairs}.  Here, we predict increases of the numbers of  levels, caused by pair-breaking, even though not necessarily visible with the applied  experimental resolution. 
In between the abrupt pair-breakings,  the number of broken pairs is almost constant and  close to integers. Saturation has been reached, and significantly more energy is needed for the next pair-breaking. 

Neutron pair-breaking dominates over  proton pair-breaking for the  energies studied. 
Even though there is a large shell gap for the protons,  
breaking of proton pairs also occurs, but then only for energies above  the proton pair-breaking energy of $2\Delta_{\rm p}$ plus the shell gap energy. According to Fig.~\ref{fig:pairs}, proton pair-breaking contributes for excitation energies above 3.5~MeV in both isotopes. 
An increased number of broken proton pairs at higher energies is expected to lead to the level density  steps at high excitation energy  being smeared out and becoming less distinctive, in accordance with the experimental findings of Sec.~\ref{subsec-steps}. 

Two effects due to the Pauli principle are notable in Fig.~\ref{fig:pairs}. In $^{119}$Sn compared to $^{118}$Sn, 1) the increases of the total average number of broken pairs  occur at higher energies; and 2) the average number of broken proton  pairs is generally higher. The explanation
probably is that  the valence neutron in $^{119}$Sn to some extent hinders the neutron pair-breaking. 
The presence of the valence  neutron makes fewer states  available for other neutrons, due to the Pauli principle. Therefore, in $^{119}$Sn compared to $^{118}$Sn, more energy is needed to break neutron pairs, and for a certain energy, proton pair-breaking  is more probable. Of course, an increase in the number of broken proton pairs leads to a corresponding decrease in the number of broken neutron pairs.

\subsection{Collective effects}

We have made use of the model to make a simple estimate of the relative impact on the level density of
 collective effects, i.e., rotations and vibrations, 
compared to that of the pair-breaking process.   
The enhancement factor of the collective effects is defined as
\begin{equation}
F_{\rm coll} (E) = \frac {\rho(E)} { \rho_{\rm non-coll}(E)}\,,
\end{equation}
where $\rho_{\rm non-coll}$ is the level density function excluding collective effects. 

\begin{figure}[tb]
\includegraphics[width=9.5cm]{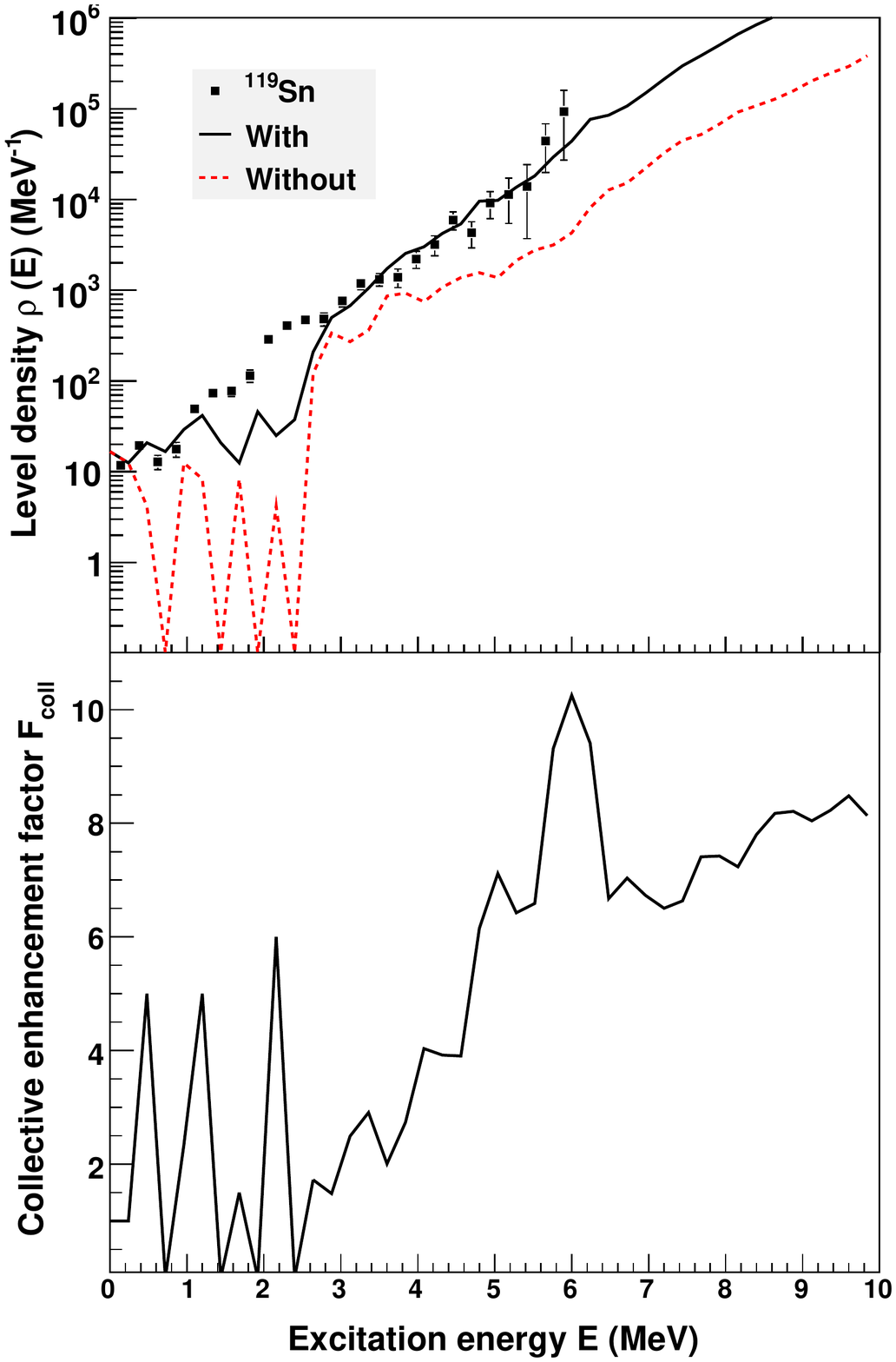}
\caption{(Color online) The impact of collective effects on the level density of $^{119}$Sn according to the combinatorial BCS model. The upper panel shows experimental level density (data points) compared with model calculations with collective effects (solid line) and without collective effects (dashed line). The lower panel shows the corresponding enhancement factor of the collective effects, $F_{\rm coll}$ (linear scale).}
\label{fig:collective}
\end{figure}

Figure \ref{fig:collective}  shows the calculated level density with and without collective contributions from vibrations and from rotational bands for $^{119}$Sn. The model prediction is assumed to be reasonably valid above $\sim 3$ MeV of excitation energy.
According to these simplistic calculations,  the enhancement factor of collective effects $F$ sharply decreases at the energies of the steps in the average number of broken quasi-particle pairs (see Fig.~\ref{fig:pairs}). For $^{119}$Sn, we find that $F$ decreases for excitation energies of approximately 2.5 and 6 MeV, where the average number of broken quasi-particle pairs increase from $\sim 0$ to $\sim 1$, and from $\sim 1$ to $\sim 2$, respectively. For the energies studied, the maximum value of $F$ is about 10, found at $E\simeq 6$ MeV. 
For $^{118}$Sn, the enhancement factor would be less than for $^{119}$Sn, since this nucleus does not have unpaired valence neutrons.

As a conclusion, the collective phenomena of vibrations and rotations  seem to have  a significantly smaller impact on the creation of new levels than the nucleon pair-breaking process, which has an enhancement factor of typically about 25 for each broken pair. 

\subsection{Parity asymmetry}
\label{combi-parity}

The parity asymmetry function $\alpha$ is defined as 
\begin{equation}
\alpha = \frac {\rho_+ - \rho_-}{\rho_+ + \rho_-}\,,
\end{equation}
where $\rho_+$ is the level density of  positive parity states, and $\rho_-$ is the level density of the negative parity states.  The values of $\alpha$ range from $-1$ to $+1$. A system with $\alpha=-1$ is obtained for $\rho_+=0$, implying that all states  have a negative parity. A system with $\alpha=0$ has equally many states with positive as negative parity and is obtained for $\rho_-=\rho_+$. 

The Nilsson scheme in Fig.~\ref{Nilsson} shows that the single-particle orbitals both above and below the neutron Fermi level are a  mixture of positive and negative  parities. In addition, each of these states may be the head of vibrational bands, for which the parity of the band may be opposite of that of the band head. 

\begin{figure}[tb]
\includegraphics[width=9.5cm]{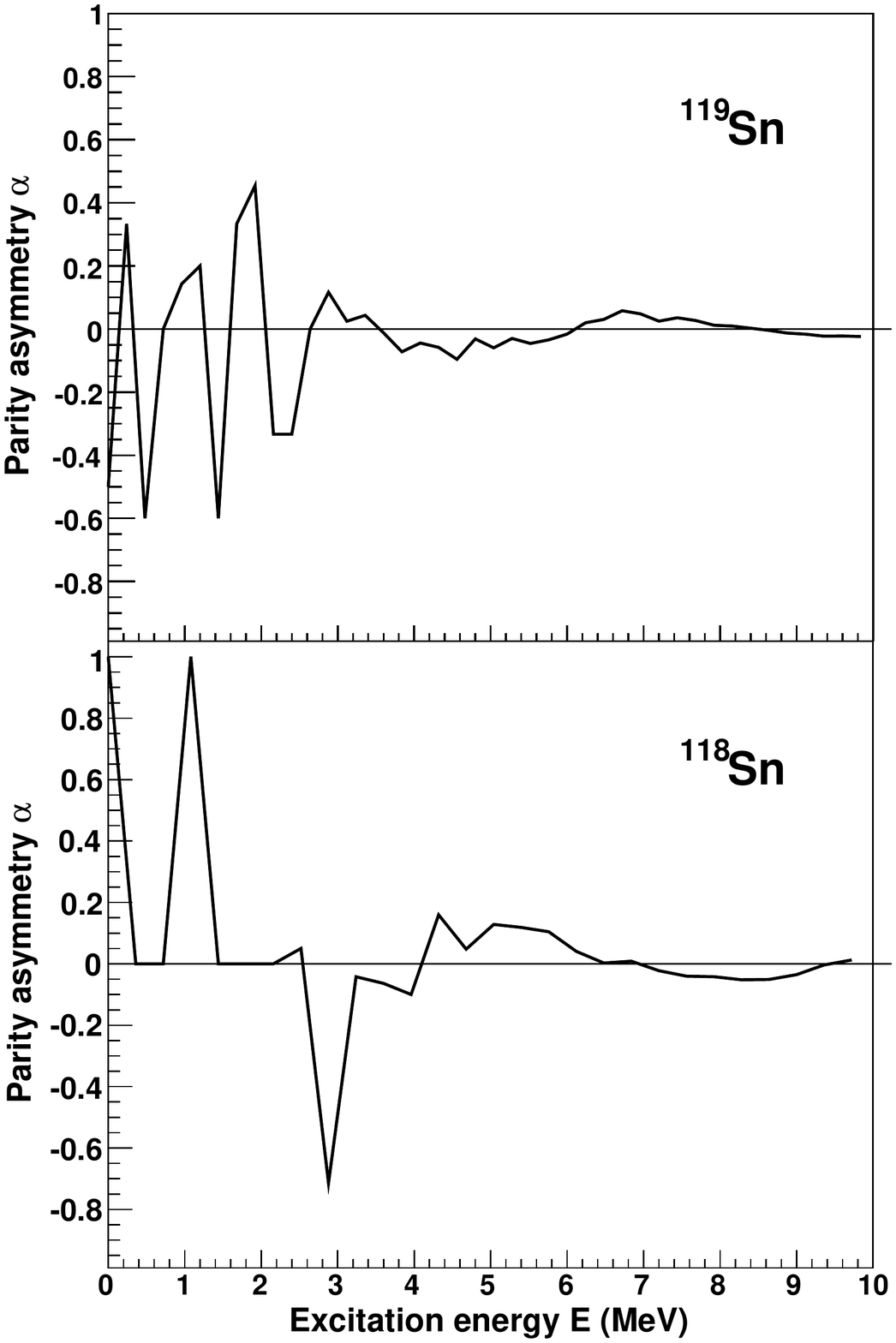}
\caption{The parity asymmetry function $\alpha$, according to the combinatorial BCS model, shown  as a function of excitation energy for $^{119}$Sn (upper panel) and  $^{118}$Sn (lower panel).}
\label{fig:parity}
\end{figure}

The parity asymmetry functions of $^{118,119}$Sn are drawn in Fig.~\ref{fig:parity}. For energies below the neutron pair-breaking energy approximately at 2.5 MeV, the even-odd isotope has a parity asymmetry function with large  fluctuations between positive and negative values, while the even-even isotope has  positive parities. This is as expected when vibrational bands of opposite parity are not introduced. (The zero parity of  $^{118}$Sn for certain low-energy regions is explained by the non-existence of energy levels.) 

Above the pair-breaking energies, the asymmetry functions begin to approach zero for both isotopes. This is also as expected, since we then have a group of  valence nucleons that will randomly occupy orbitals of positive and negative parity and  on average give an $\alpha$ close to zero. Above $\sim 4$~MeV, the parity distributions of $^{118,119}$Sn are symmetric with $\rho_+ \simeq \rho_-$. This is a gratifying  property, since parity symmetry  is an assumption in the Oslo method normalization procedure for both the level density and the $\gamma$-ray strength function (see Sec.~\ref{section-gamma}).

\subsection{Spin distribution}

The combinatorial BCS model determines the total spin $I$ for each level from the relation:
\begin{equation}
\label{eqn:model-spin}
I\left(I+1\right)=R\left(R+1\right)+ \sum_{\Omega_\pi,\, \Omega_\nu} \Omega_\pi + \Omega_\nu \,,
\end{equation}
from which the  spin distribution of the level density $\rho$ may be estimated. 

\begin{figure*}[h!]
\includegraphics[height=14cm]{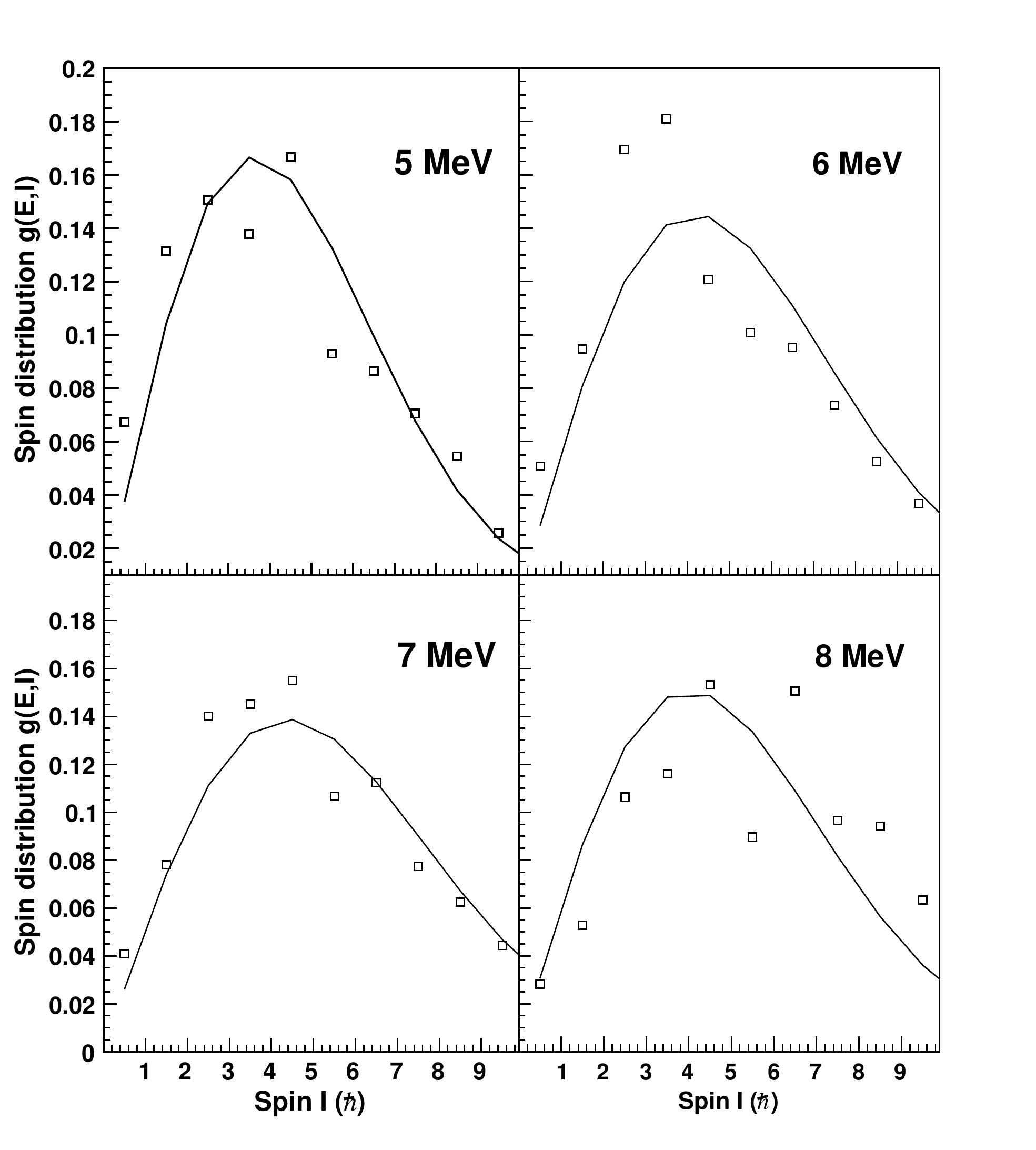}
\caption{Comparison of spin distributions in $^{118}$Sn at different excitation energies. The open squares are calculated from the combinatorial BCS model in Eq.~(\ref{eqn:model-spin}), while the solid lines are the predictions of Gilbert and Cameron in Eq.~(\ref{eqn:GC-spin}). The spin distributions are normalized to 1.} 
\label{fig:spindis}
\end{figure*}

The resulting spin distribution  is compared with the theoretical spin distribution of Gilbert and Cameron in Eq.~(\ref{eqn:GC-spin}), using the same parameterization of the spin cutoff parameter as Eq.~(\ref{eqn:cutoff}). Figure \ref{fig:spindis} shows the comparison for four different excitation energies: 5, 6, 7 and 8 MeV. The agreement is generally good.
Hence, the spin calculation in Eq.~(\ref{eqn:model-spin}), and the assumption of the rigid rotational parameter $A_{\rm rigid}$ obtained at  $S_n$, are indicated  to be  reasonable assumptions. We also note from the figure that the average spin, $\left< I(E) \right>$, is only slowly increasing with excitation energy, justifying the pseudo entropy definition introduced in Eq.~(\ref{eq:pseudoentropi}).

\section{Gamma-ray strength functions}

\label{section-gamma}

\subsection{Normalization and experimental results}
\label{sub-norm}

The  $\gamma$-ray transmission coefficient ${\cal T}$, which is deduced from the experimental data, is related to the
$\gamma$-ray strength function $f$  by
\begin{equation}
{\cal T}(E_\gamma) = 2\pi \sum_{XL}E_\gamma^{2L+1}f_{XL}(E_\gamma)\,,
\end{equation}
where $X$ denotes the electromagnetic character and $L$ the multipolariy of the  $\gamma$-ray.
The transmission coefficient ${\cal T}$  is normalized in slope  $\alpha$  and in absolute value $B$ according to Eq.~(\ref{eq:array2}). The slope was determined in Sec.~\ref{LD} in the case of $^{118,119}$Sn, and in Ref.~\cite{Sn_Density} in the case of $^{116}$Sn. The absolute value normalization is yet to be determined. This is done using the literature values of the
average total radiative width at the neutron separation energy,
$\left<\Gamma_\gamma(S_n)\right>$, which are measured for neutron capture reactions $(n,\gamma)$. 

The $\gamma$-ray transmission coefficient ${\cal T}$ is related to the  average total radiative width $\left<  \Gamma_\gamma (E, I, \pi)\right>$ of levels  with energy $E$, spin $I$ and parity $\pi$ by \cite{Kopecky}:
\begin{eqnarray}
\label{eq:generalexpression}
\lefteqn { \left<  \Gamma_\gamma (E, I, \pi)\right> =  \frac {1}{2\pi \, \rho(E,I,\pi)} } \nonumber \\
& &  \times \sum_{XL} \sum_{I_f,\, \pi_f} \int_{E_\gamma}^E {\rm d}E_\gamma {\cal T}_{XL}(E_\gamma) \, \rho(E-E_\gamma,I_f,\pi_f) \,.
\end{eqnarray}
The summations and integration are over all final levels of spin $I_f$ and parity $\pi_f$ that are accessible through a $\gamma$-ray transition categorized by the energy $E_\gamma$, electromagnetic character $X$ and multipolarity $L$. 

For  $s$-wave neutron resonances and assuming a major contribution from dipole radiation and
 parity symmetry for all excitation energies,
 the general expression in Eq.~(\ref{eq:generalexpression}) will at $S_n$ reduce to
\begin{eqnarray}
\label{eq:Gammagamma}
\lefteqn { \left<  \Gamma_\gamma (S_n, I_t \pm 1/2,\pi_t)\right> } \nonumber \\
& = & \frac {B}{4\pi\,\rho(S_n, I_t \pm 1/2, \pi_t )} \int_0^{S_n} {\rm d}E_\gamma \, {\cal T}(E_\gamma)\,\rho(S_n-E_\gamma) \nonumber \\
& & \times \sum_{J = -1}^{1}g (S_n - E_\gamma, I_t \pm 1/2 + J)\,.
\end{eqnarray}
Here, $I_t$ and $\pi_t$ are the spin and parity of the target nucleus in the  $(n,\gamma)$ reaction.
Indeed, the results from the combinatorial BCS model in  Sec.~\ref{sec:combi}   supports the symmetry assumption of the  parity distribution.
The normalization constant  $B$  in Eq.~(\ref{eq:Gammagamma}) is determined \cite{Voinov01} by replacing ${\cal T}$ with
the experimental transmission coefficient, $\rho$ with
the experimental level density, $g$ with the spin distribution given in Eq.~(\ref{eqn:GC-spin}), and $\left<  \Gamma_\gamma(S_n)\right>$ with its
literature value. 

The input parameters needed for  determining the normalization constant $B$ for $^{118,119}$Sn  are  shown in Tab.~\ref{tab:styrke-normering} and taken from Ref.~\cite{RIPL-3}. 
For  $^{116}$Sn,  the level spacing $D_0(S_n)$  is not available in the literature. Therefore, $\rho(S_n)$ was estimated from systematics for the normalization of $\alpha$
in Ref.~\cite{Sn_Density}. The value of $D_0$  in Tab.~\ref{tab:styrke-normering}  is  estimated from $\rho(S_n)$.
Note that there was an error in the spin cutoff parameters $\sigma(S_n)$ of $^{116,117}$Sn in Refs.~\cite{Sn_Density, Sn_Strength}. 
The impact of this correction on the normalization of level densities and strength functions is very small. Moreover, updated values of $D_0(S_n)$ and $\left<\Gamma_\gamma(S_n)\right>$ are now available for  $^{117}$Sn \cite{RIPL-3}.  All the new normalization parameters for $^{116,117}$Sn are presented in Tab.~\ref{tab:oppdatert}. The value of $\left<\Gamma_\gamma(S_n)\right>$ of  $^{116}$Sn is taken from the indicated value  in Ref.~\cite{RIPL-3}.
\begin{table}[!htb] 
\caption{Input parameters for normalization of the $\gamma$-ray transmission coefficient ${\cal T}$  for $^{118,119}$Sn.}
\begin{tabular}{|l|ccc|}
\hline
\hline
Nucleus   & $I_t$ & $D_0(S_n)$ & $\left<\Gamma_{\gamma}(S_n)\right>$  \\ 
       &  ($\hbar$) & (eV) & (meV) \\
\hline
$^{119}$Sn  &  0 & 700 & 45 \\
$^{118}$Sn   & 1/2 & 61 & 117 \\
\hline
\hline
\end{tabular}
\label{tab:styrke-normering}
\end{table}
\begin{table}[!htb] 
\caption{New normalization parameters for $^{116,117}$Sn.}
\begin{tabular}{|l|cccc|c|}
\hline
\hline
Nucleus   & $\sigma(S_n)$ & $D_0(S_n)$ & $\rho(S_n)$ & $\left<\Gamma_\gamma(S_n)\right>$ & $\eta$  \\ 
       &   & (eV) & ($10^4$ MeV$^{-1}$) & (meV) & \\
\hline
$^{117}$Sn   & 4.58 & 450(50) & 9.09(2.68) & 52 &  0.43 \\
$^{116}$Sn   & 4.76 &  59 & 40(20) & 120 & 0.45 \\
\hline
\hline
\end{tabular}
\label{tab:oppdatert}
\end{table}

\begin{figure*}[tb]
\includegraphics[height=13.4cm]{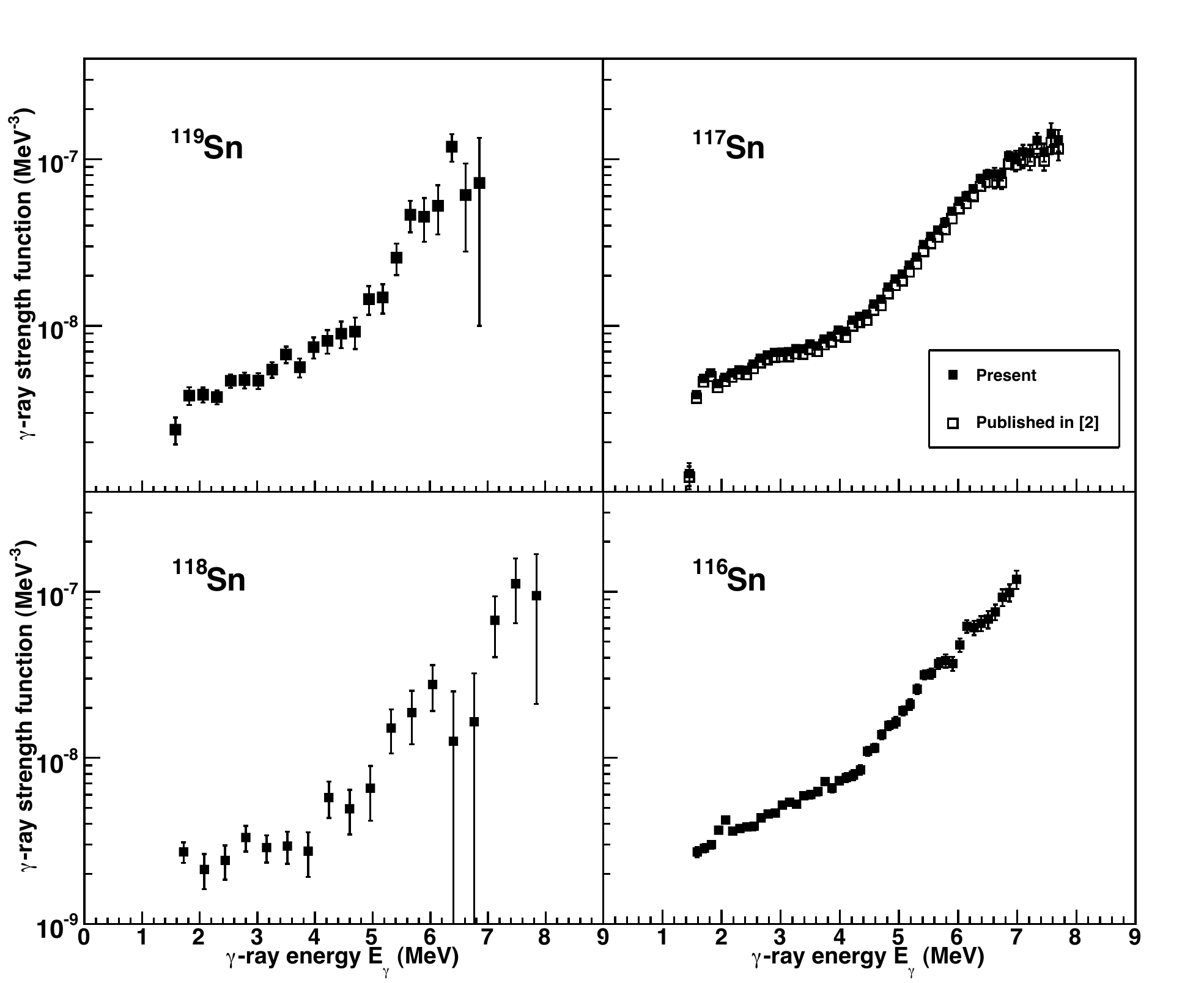}
\caption{Normalized $\gamma$-ray strength functions as functions of $\gamma$-ray energy. The upper panels show $^{119}$Sn (left) and both present and  previously published versions of
 $^{117}$Sn  (right). The energy bins  120 and 240 keV/ch for $^{117,119}$Sn, respectively. 
 The lower panels show $^{118}$Sn (left) and $^{116}$Sn (right)  with the energy bins of 360 and 120 keV/ch, respectively.}
\label{fig:strength_both}
\end{figure*}

The resulting $\gamma$-ray strength functions of  $^{116-119}$Sn are shown in Fig.~\ref{fig:strength_both}.  For all isotopes, it is clear that there is a change of the log-scale derivate at $E_\gamma \sim 4 $ MeV, leading to a sudden increase of strength.  Such an increase may indicate the onset of a resonance. 
The comparison  in Fig.~\ref{fig:strength_both}  of the 
 new $^{117}$Sn strength function with the earlier published one \cite{Sn_Strength} confirms that
 correcting the 
 $\sigma(S_n)$ and the $D_0(S_n)$ had only a minor impact on the normalization.  

\begin{figure}[tb]
\includegraphics[width=9.5cm]{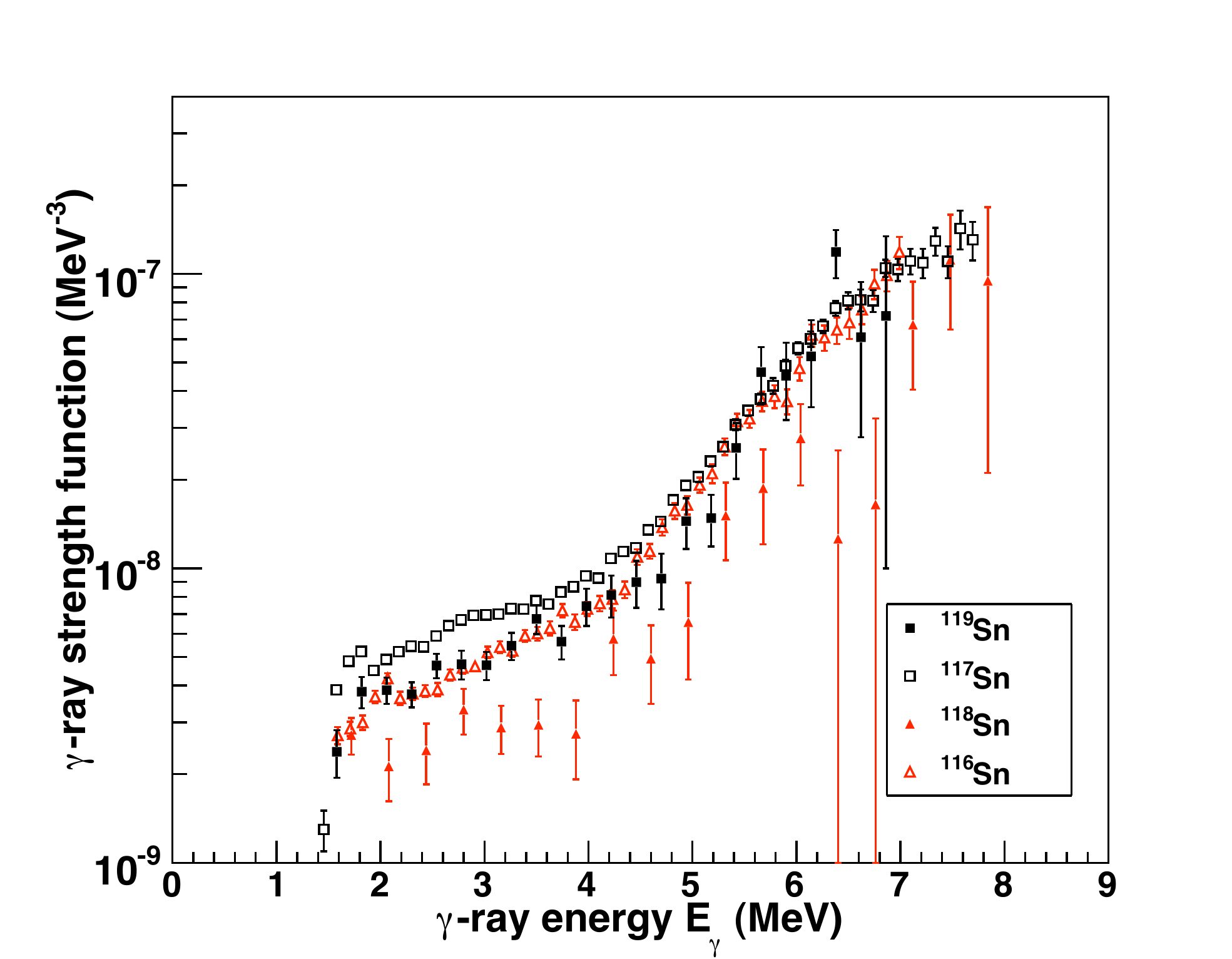}
\caption{(Color online) The four normalized strength functions of $^{116-119}$Sn shown together.}
\label{fig:alle4}
\end{figure}

Figure \ref{fig:alle4} shows the four normalized strength functions of $^{116-119}$Sn  together. 
They are all approximately equal except for $^{118}$Sn, which has a lower absolute normalization than the others. This  is surprising considering the quadrupole deformation parameter of $^{118}$Sn ($\varepsilon_2=0.111$) being almost identical to that of $^{116}$Sn ($\varepsilon_2=0.112$)  \cite{RIPL-2}. 
In the following, we therefore multiply the strength of $^{118}$Sn with a factor of 1.8 to get it on the same footing as the others.
The values of $\rho(S_n)$ and the scaling parameter $\eta$ (see also Sec.~\ref{LD-results}) of these Sn isotopes are collected in Tab.~\ref{tab:eta}. 
For $^{118}$Sn, the $\rho(S_n)$  may be expected  to be larger, while  $\eta$  may be expected to be smaller. 
It would be desirable to remeasure both $D_0(S_n)$ and $\left<\Gamma_\gamma\right>$ for this isotope, since the apparent wrong normalization of the strength function of $^{118}$Sn depends on these parameters.

\begin{table}[!htb] 
\caption{The Fermigas approximation for $\rho(S_n)_{\rm BSFG}$, the calculated $\rho(S_n)$, and the resulting scaling parameter $\eta$, for $^{116-119}$Sn.}
\begin{tabular}{|l|cc|c|}
\hline
\hline
Nucleus   & $\rho(S_n)_{\rm BSFG}$ & $\rho(S_n)$ & $\eta$    \\
  & ($10^4$ MeV$^{-1}$) &  ($10^4$ MeV$^{-1}$) & \\
\hline
$^{119}$Sn  & 14 & 6.05(175) & 0.44 \\
$^{118}$Sn   &  65 & 38.4(86) &  0.59 \\
$^{117}$Sn   &  22 &9.09(2.68) & 0.43\\
$^{116}$Sn   &   89 & 40.0(20.0) & 0.45 \\
\hline
\hline
\end{tabular}
\label{tab:eta}
\end{table}

\subsection{Pygmy resonance}

Comparing our measurements with other experimental data makes potential resonances easier to localize.
Experimental cross section data $\sigma(E_\gamma)$
are converted to  $\gamma$-ray strength $f(E_\gamma)$ through the relation:
\begin{equation}
f(E_\gamma) = \frac{1}{3\pi \hbar^2 c^2}\left(\frac{\sigma(E_\gamma)}{E_\gamma}\right)\,.
\end{equation}

\begin{figure*}[tb]
\includegraphics[height=13.4cm]{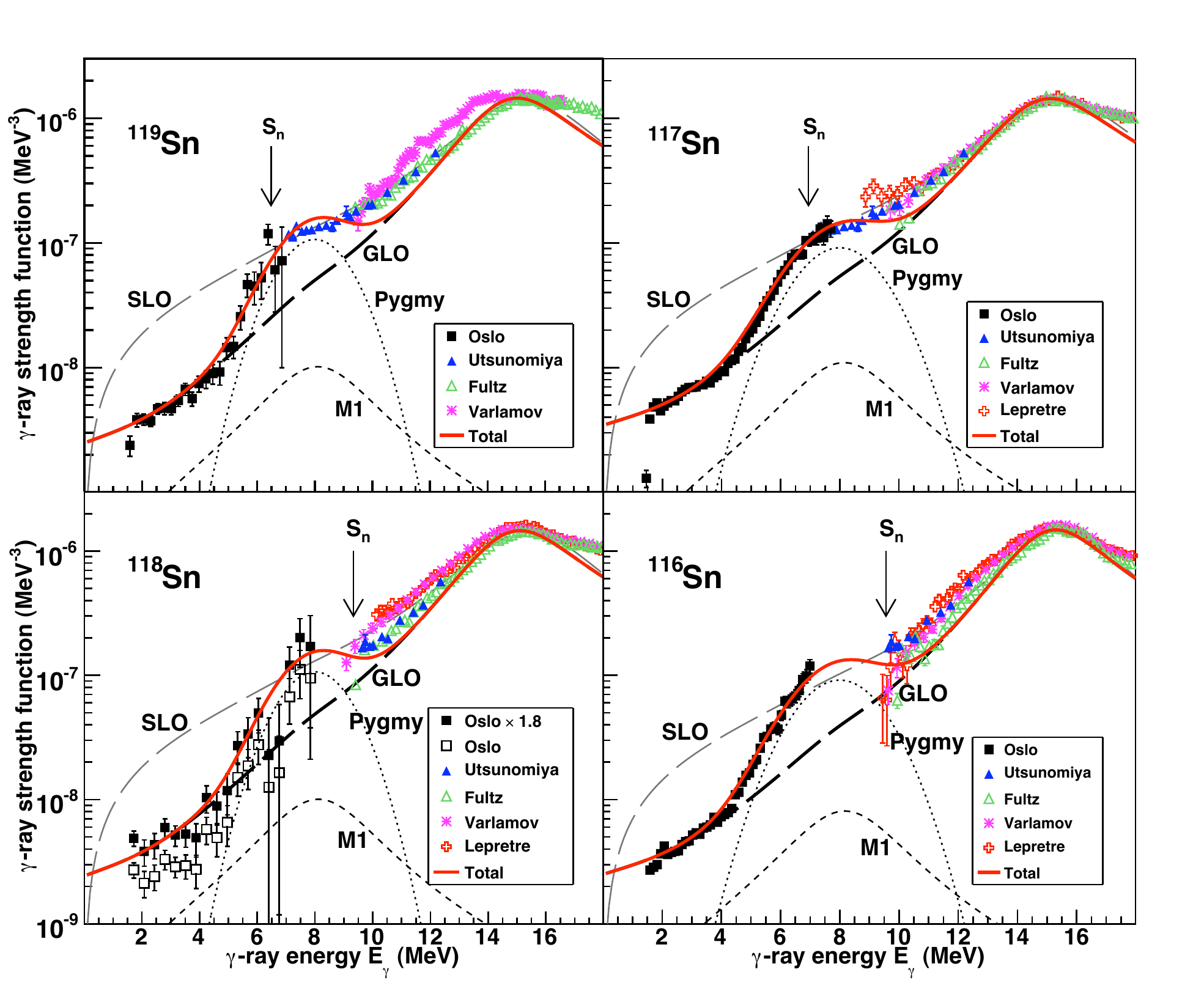}
\caption{(Color online) Comparison of theoretical predictions including pygmy fits  with experimental measurements for  $^{116-119}$Sn.  The total strengths (solid lines) are modeled as Gaussian pygmy additions to the GLO (E1 + M1) baselines. The SLO (E1 + M1) baselines are also shown, failing  to  reproduce the measurements for low $E_\gamma$. The arrows indicate the neutron separation energies $S_n$.\\
Upper, left panel: Comparison of theoretical predictions of  $^{119}$Sn with the Oslo measurements, $^{117}{\rm Sn}(\gamma,n)$ from Utsunomiya {\em et al.}~\cite{Utsunomiya}, $^{119}{\rm Sn}(\gamma,x)$ from Fultz {\em et al.}~\cite{Fultz69}, and $^{119}{\rm Sn}(\gamma,n)$ from Varlamov {\em et al.}~\cite{Varlamov09}.\\
Upper, right panel: 
Comparison of theoretical predictions of  $^{117}$Sn with the Oslo measurements, $^{117}{\rm Sn}(\gamma,n)$ from Utsunomiya {\em et al.}~\cite{Utsunomiya},  $^{117}{\rm Sn}(\gamma,x)$ from Fultz {\em et al.}~\cite{Fultz69}, $^{117}{\rm Sn}(\gamma,x)$ from Varlamov {\em et al.}~\cite{Varlamov03}, and $^{117}{\rm Sn}(\gamma,x)$ from Lepr\^{e}tre {\em et al.}~\cite{Lepretre74}. \\
Lower, left panel: Comparison of theoretical predictions of  $^{118}$Sn with the Oslo measurements multiplied with 1.8 (filled squares) (the measurements with the original normalization are also included as open squares), $^{116}{\rm Sn}(\gamma,n)$ from Utsunomiya {\em et al.}~\cite{Utsunomiya},  $^{118}{\rm Sn}(\gamma,x)$ from Fultz {\em et al.}~\cite{Fultz69}, $^{118}{\rm Sn}(\gamma,x)$ from Varlamov {\em et al.}~\cite{Varlamov03}, and $^{118}{\rm Sn}(\gamma,x)$ from Lepr\^{e}tre {\em et al.}~\cite{Lepretre74}.\\
Lower, right panel: Comparison of theoretical predictions of  $^{116}$Sn with the Oslo measurements, $^{116}{\rm Sn}(\gamma,n)$ from Utsunomiya {\em et al.}~\cite{Utsunomiya},  $^{116}{\rm Sn}(\gamma,x)$ from Fultz {\em et al.}~\cite{Fultz69}, $^{116}{\rm Sn}(\gamma,x)$ from Varlamov {\em et al.}~\cite{Varlamov03}, and $^{116}{\rm Sn}(\gamma,x)$ from Lepr\^{e}tre {\em et al.}~\cite{Lepretre74}.
}
\label{fig:pygme}
\end{figure*}

Figure \ref{fig:pygme} shows the comparison of the Oslo strength functions of $^{116-119}$Sn with those of the photoneutron cross section reactions $^{116,117}{\rm Sn}(\gamma,n)$ from Utsunomiya {\em et al.}~\cite{Utsunomiya} and
$^{119}{\rm Sn}(\gamma,n)$ from Varlamov {\em et al.}~\cite{Varlamov09},
photoabsorption reactions $^{116-119}{\rm Sn}(\gamma,x)$ from Fultz {\em et al.}~\cite{Fultz69}, 
$^{116-118}{\rm Sn}(\gamma,x)$ from Varlamov {\em et al.}~\cite{Varlamov03}, 
and $^{116-118}{\rm Sn}(\gamma,x)$ from Lepr\^{e}tre {\em et al.}~\cite{Lepretre74}.
Clearly,  the measurements on $^{117,119}$Sn both from Oslo and from Utsunomiya {\em et al.}~\cite{Utsunomiya}   independently indicate a resonance, from the changes of slopes. 
For $^{116,118}$Sn, the Oslo data clearly shows  the presence of resonances.
Hence, the resonance earlier observed in $^{117}$Sn \cite{Sn_Strength} is confirmed also in $^{116,118,119}$Sn.
This resonance will be referred to as the pygmy resonance.

In order to investigate the experimental strength functions further, and in particular the pygmies, we have applied commonly used models for the Giant Electric Dipole Resonance (GEDR) and for the magnetic spin-flip resonance, also known as the Giant Magnetic Dipole Resonance (GMDR).

For the GEDR resonance, the Generalized Lorentzian (GLO) model \cite{Kopecky87} is used. The GLO  model is known to  agree well both for  low $\gamma$-ray energies, where we measure, and for the 
GEDR centroid 
at about 16 MeV. The strength function approaching a non-zero value for low $E_\gamma$ 
is not a property specific for the Sn isotopes, but has  been the case for all nuclei studied at the OCL so far. 

In the GLO model, the $E1$ strength function is given by  \cite{Kopecky87}:
\begin{eqnarray}
f^{\rm GLO}_{E1} (E_\gamma)  &=&  \frac{1}{3\pi^2\hbar^2 c^2}\sigma_{E1}\,\Gamma_{E1} \nonumber \\
& \times &\left[ E_\gamma \,\frac{\Gamma_{KMF}(E_\gamma,T_f)}{\left({E_\gamma}^2 - {E_{E1}}^2\right)^2 + {E_\gamma}^2 \left(\Gamma_{KMF}(E_\gamma,T_f)\right)^2} \right. \nonumber \\
  &+& \left. 0.7 \, \frac{\Gamma_{KMF}(E_\gamma=0, T_f)}{{E_{E1}}^3}\right]  \nonumber \\
 \end{eqnarray}
in units of ${\rm MeV}^{-3}$, 
where the Lorentzian parameters are the GEDR's centroid energy $E_{E1}$, width $\Gamma_{E1}$ and cross section $\sigma_{E1}.$
We use the experimental parameters of Fultz  \cite{Fultz69}, shown in Tab.~\ref{tab:parametre-teori}.
 The GLO model is  
temperature dependent from the incorporation of a temperature dependent width $\Gamma_{KMF}$. This width    is   the term responsible  for  ensuring  the non-vanishing GEDR at low excitation energy. It has been adopted from the Kadmenski{\u{\i}}, Markushev and Furman (KMF) model \cite{KMF} and is given by:
\begin{equation}
\Gamma_{KMF}(E_\gamma, T_f )= \frac{\Gamma_r}{{E_r}^2}\left( {E_\gamma}^2 + 4\pi^2 {T_f}^2 \right)\,,
\end{equation}
in units of MeV.

Usually,  $T_f$  is interpreted as the nuclear temperature of the final state, with the commonly applied expression $T_f=~\sqrt{U/a}$. 
On the other hand, we are assuming a constant temperature, i.e., the $\gamma$-ray strength function is  
independent of excitation energy. This approach  is adopted for consistency with the Brink-Axel hypothesis (see Sec.~\ref{sec:setup}), where the strength function was assumed to be temperature independent.
Moreover, we treat $T_f$  as a free parameter 
in order to fit   in  the best possible way  the theoretical strength prediction to    the  low energy measurements. The applied values of $T_f$ are listed in Tab.~\ref{tab:parametre-teori}. 

\begin{table}[!htb] 
\caption{Parameters used for the theoretical $\gamma$-ray strength functions of $^{116-119}$Sn. The value of $T_f$ in $^{118}$Sn has been found for the  measured strength function multiplied by 1.8.}
\begin{tabular}{|l|ccccccc|}
\hline
\hline
Nucleus   & $E_{E1}$ &  $\Gamma_{E1}$ & $\sigma_{E1} $ & $E_{M1}$ & $\Gamma_{M1}$ & $\sigma_{M1}$  & $T_f$ \\ 
       &  (MeV) &  (MeV) & (mb)& (MeV) &  (MeV) & (mb) & (MeV) \\
\hline
$^{119}$Sn  &  15.53 &  4.81 & 253.0 & 8.34 & 4.00 & 0.963 &  0.40(1)  \\
$^{118}$Sn   &  15.59 &  4.77 & 256.0 & 8.36 & 4.00 & 0.956  & 0.40(1) \\
$^{117}$Sn  &   15.66 & 5.02   & 254.0 &  8.38 & 4.00 & 1.04 &  0.46(1)  \\
$^{116}$Sn   &  15.68 &  4.19 & 266.0 &  8.41 & 4.00 & 0.773  & 0.46(1) \\
\hline
\hline
\end{tabular}
\label{tab:parametre-teori}
\end{table}

The $M1$ spin-flip resonance is modeled with the functional form of  a Standard Lorentzian (SLO)  \cite{RIPL-2}:
\begin{equation}
f_{M1}^{\rm SLO}(E_\gamma) = \frac{1}{3\pi^2\hbar^2 c^2} \frac{\sigma_{M1} {{\Gamma}_{M1}}^2 E_\gamma}{\left({E_\gamma}^2 - {E_{M1}}^2\right)^2 + {E_\gamma}^2 \, {\Gamma_{M1}}^2} \,,
\end{equation}
where the parameter $E_{M1}$ is the centroid energy, $\Gamma_{M1}$ the width and $\sigma_{M1}$ the cross section, of the GMDR. These Lorentzian parameters are predicted from the expressions in Ref.~\cite{RIPL-2}, with the results as shown  in Tab.~\ref{tab:parametre-teori}.

In the absence of any established theoretical prediction about the pygmy resonance, we found that the pygmy is satisfactorily   reproduced by a Gaussian distribution \cite{Sn_Strength}:
\begin{equation}
f_{\rm pyg}(E_\gamma) = C_{\rm pyg} \cdot \frac{1}{\sqrt{2\pi} \, \sigma_{\rm pyg}} \exp\left[ \frac{-\left( E_\gamma - E_{\rm pyg}\right)^2}{2{\sigma_{\rm pyg}}^2}\right) \,,
\end{equation}
where  $C_{\rm pyg}$ is the pygmy's normalization constant, $E_{\rm pyg}$ the energy centroid and  $\sigma_{\rm pyg}$  the standard deviation.  These parameters are treated as  free. The  total model prediction of the $\gamma$-ray strength function is then given by:
\begin{equation}
f_{\rm total} = f_{E1} + f_{M1} + f_{\rm pyg}\,.
\end{equation}

By adjusting the Gaussian pygmy parameters to make the best fit to the experimental data of $^{116-119}$Sn, we obtained the values as presented in Tab.~\ref{tab:pygme}.  
The pygmy fit of $^{117}$Sn is updated  and corresponds to the present normalization of the strength function. The pygmy fit 
also gave an excellent fit  for $^{116}$Sn. For $^{118,119}$Sn, it was necessary to  slightly reduce the values of   $T_f$ and $\sigma_{\rm pyg}$. The similarity of the sets of parameters for the different nuclei is gratifying. 
As is seen in Fig.~\ref{fig:pygme}, these theoretical predictions describe the measurements rather well.

\begin{table}[!htb] 
\caption{Empirical values of $^{116-119}$Sn pygmies  Gaussian  parameters, the   integrated  pygmy strengths  and the TRK values. For $^{118}$Sn, the values have been found from fitting to the measured strength function multiplied by 1.8.}
\begin{tabular}{|l|ccc|cc|}
\hline
\hline
Nucleus   & $C_{\rm pyg}$ & $E_{\rm pyg}$  & $\sigma_{\rm pyg }$ & Int. strength &TRK value  \\ 
       & ($10^{-7}{\rm MeV}^{-2})$ & (MeV) & (MeV)  & (MeV$\cdot$mb) & (\%) \\
\hline
$^{119}$Sn  & 3.2(3) & 8.0(1)  & 1.2(1) & 30(15) & 1.7(9) \\
$^{118}$Sn   & 3.2(3)  &  8.0(1) & 1.2(1) & 30(15) & 1.7(9) \\
$^{117}$Sn  & 3.2(3) & 8.0(1)  & 1.4(1) & 30(15) & 1.7(9) \\
$^{116}$Sn & 3.2(3) & 8.0(1) & 1.4(1)  & 30(15) & 1.7(9) \\
\hline
\hline
\end{tabular}
\label{tab:pygme}
\end{table}

The pygmy centroids of  all the isotopes are estimated to be around 8.0(1) MeV. It is noted that an earlier experiment  by Winhold {\em et al.}~\cite{Winhold} using the $(\gamma,n)$ reactions determined the pygmy centroids for $^{117,119}$Sn to approximately 7.8 MeV, in agreement with our measurements. 
 
 Extra strength  has been added in the energy region of $\sim(4 - 11)$~MeV. 
 The total  integrated pygmy strengths are 30(15)~MeV$\cdot$mb for all four isotopes. This constitutes $1.7(9)\%$ 
 of the classical Thomas-Reiche-Kuhn (TRK) sum rule, assuming all pygmy strength being $E1$. 
 Even though these resonances are rather small compared to the GEDR, they may have a non-negligible impact on  nucleosynthesis in supernovas \cite{Goriely98}.

If one does not multiply the strength
function of $^{118}$Sn by 1.8 for the footing equality, then the pygmy of  $^{118}$Sn becomes very different from those of the other isotopes, and the total prediction is not able to  follow as well the measurements  for low $E_\gamma$. A pygmy fit of the original normalization does however give: $T_f= 0.28(2)$ MeV, $C_{\rm pyg}=1.8(6)\cdot 10^{-7} {\rm MeV}^{-2}$, $E_{\rm pyg} = 8.0(2)$ MeV and $\sigma_{\rm pyg} =1.0(1)$ MeV. This represents a smaller pygmy, giving an integrated strength of 17(8)~MeV$\cdot$mb and a  TRK value of $1.0(5)\%$.

The commonly applied Standard Lorentzian (SLO) was also tested as a model of the  baseline and is included in Fig.~\ref{fig:pygme}. The SLO succeeds in reproducing the $(\gamma,x)$ data, but
clearly fails for the low-energy strength measurements, both when it comes to   absolute value and to shape.
The same has been the case also for many other nuclei measured at the OCL. Therefore, we deem the SLO not to be adequate below the neutron threshold. 

Probably, the pygmies of all the Sn isotopes are caused by the same phenomenon. It is still indefinite whether the Sn pygmy is of $E1$ or $M1$ character.
 A clarification  would be of utmost importance. 

Earlier studies indicate an $E1$ character of the Sn pygmy. Amongst these are the nuclear resonance fluorescence experiments (NRF) performed on  $^{116,124}$Sn \cite{Govaert}   and on $^{112,124}$Sn \cite{Tonchev}, and the Coulomb dissociation  experiments performed on $^{129-132}$Sn \cite{Adrich,Klimkiewicz}.
If the Sn pygmy is of $E1$ character, it may be consistent  with the so-called $E1$ neutron skin oscillation mode, discussed in Refs.~\cite{Paar, Sarchi, Isacker}.

However, the possibility of an $M1$ character cannot be ruled out. Figure \ref{Nilsson} shows that the Sn isotopes have their proton Fermi level located right in between the $g_{7/2}$ and $g_{9/2}$ orbitals,
and their  neutron Fermi level between $h_{11/2}$ and $h_{9/2}$. 
Thus,  an enhanced $M1$ resonance may be due to  proton $g_{7/2} \leftrightarrow g_{9/2}$ and neutron $h_{11/2} \leftrightarrow h_{9/2}$  magnetic spin-flip transitions. The existence of an $M1$ resonance in this energy region has been indicated in an earlier experimental study: proton inelastic scattering experiment on  $^{120,124}$Sn  \cite{Djalali}. 

\section{Conclusions}  
The level densities of $^{118,119}$Sn and the $\gamma$-ray strength functions of $^{116,118,119}$Sn have been measured using the
($^3$He, \,$\alpha \gamma$) and ($^3$He,$^3$He$^\prime\gamma$) reactions and the Oslo method. 

The level density function of $^{119}$Sn shows pronounced steps for excitation energies below $\sim 4$ MeV. 
This may be explained by the fact that Sn has a closed proton shell, so that only neutron pairs are broken at low energy. Without any proton pair-breaking smearing out the level density function, the steps from neutron pair-breaking remain distinctive.
The entropy has been deduced from the experimental level density functions, with a mean value of the single neutron entropy in $^{119}$Sn determined to $(1.7 \pm 0.2) \, k_B$. These findings are in good agreement with those of $^{116,117}$Sn.

A combinatorial BCS model has been used to extract nuclear properties from the experimental level density. 
The number of broken  proton and neutron pairs as a function of excitation energy is deduced, showing that neutron pair-breaking is the most dominant pair-breaking process for the entire energy region studied. 
The enhancement factor of collective effects on  level density  contributes a maximum factor of about 10, which is small compared to that of pair-breaking. The parity distributions are found to be symmetric above $\sim4$~MeV of excitation energy.

In all the  $^{116-119}$Sn strength functions, a significant enhancement  is observed in the energy region of $E_\gamma \simeq (4-11)$ MeV. The  integrated strength of the resonances  correspond to $1.7(9)\%$
 of the TRK sum rule. 
These  findings are in agreement with the conclusions of earlier studies. 

\acknowledgments 
The authors wish to thank E.~A.~Olsen, J.~Wikne and A.~Semchenkov for excellent experimental conditions.
The funding of this research from  The Research Council of Norway (Norges forskningsråd) is gratefully acknowledged. This work was supported in part by the U.~S.~Department of Energy grants No.~DE-FG52-09-NA29640 and No.~DE-FG02-97-ER41042.


\begin{references} 

\bibitem{Sn_Density} U. Agvaanluvsan,  A.~C.~Larsen, M.~Guttormsen, R.~Chankova, G.~E.~Mitchell, A.~Schiller,  S.~Siem, and A.~Voinov, Phys.~Rev. C {\bf 79}, 014320 (2009).

\bibitem{Sn_Strength} U.~Agvaanluvsan, A.~C.~Larsen, R.~Chankova, M.~Guttormsen, G.~E.~Mitchell, A.~Schiller,  S.~Siem, and A.~Voinov, Phys. Rev. Lett. {\bf 102}, 162504 (2009).

\bibitem{Gut96}M.~Guttormsen, T.~S.~Tveter, L.~Bergholt, F.~Ingebretsen, and J.~Rekstad, Nucl.\ Instrum.\ Methods Phys.\ Res.\ A \textbf{374}, 371 (1996). 

\bibitem{Gut87} M.~Guttormsen, T.~Rams{\o}y, and J.~Rekstad, Nucl.\ Instrum.\ Methods Phys.\ Res.\ A {\bf 255}, 518 (1987).

\bibitem{Sch00a}A.~Schiller, L.~Bergholt, M.~Guttormsen, E.~Melby, J.~Rekstad, and S.~Siem, Nucl.\ Instrum.\ Methods Phys.\ Res.\ A \textbf{447}, 498 (2000).  

\bibitem{Bohr-Mottelson} A. Bohr and B. Mottelson, Nuclear Structure (Benjamin, New York, 1969), Vol. I. 

\bibitem{Bri55} D.~M.~Brink, Ph.D. thesis, Oxford University, 1955.

\bibitem{Axe62} P.~Axel, Phys. Rev. \textbf{126}, 671 (1962).

\bibitem{Mughabghab} S.~F.~Mughabghab, Atlas of Neutron Resonances, Fifth Edition, Elsevier Science (2006).

\bibitem{RIPL-3} R. Capote {\em et al.}, 
{\em RIPL-3 – Reference Input Parameter Library for Calculation of Nuclear Reactions and Nuclear Data Evaluations}, 
Nuclear Data Sheets {\bf 110} (2009), 3107-3214. Available online at http://www-nds.iaea.org/RIPL-3/.

\bibitem{Egi88} T.~von~Egidy, H.~H.~Schmidt, and A.~N.~Behkami, Nucl.\ Phys.\ \textbf{A481}, 189 (1988). 

\bibitem{Egidy} T.~von~Egidy and D.~Bucurescu, Phys.~Rev.\ C {\bf 72}, 044311 (2005); {\bf 73}, 049901(E) (2006).

\bibitem{Utsunomiya} H. Utsunomiya, S. Goriely, M. Kamata, T. Kondo, O. Itoh, H. Akimune, T. Yamagata, H. Toyokawa, Y.-W. Lui, S. Hilaire, and A. J. Koning, Phys.\ Rev.\ C \bf 80\rm, 055806 (2009).

\bibitem{Gil65} A.~Gilbert and A.~G.~W.~Cameron, Can.\ J.\ Phys.\ {\bf 43}, 1446 (1965).

\bibitem{Wapstra} G.~Audi and A.~H.~Wapstra, Nucl. Phys. {\bf A595}, 409 (1995).

\bibitem{Dob01} J.~Dobaczewski, P.~Magierski, W.~Nazarewicz, W.~Satu{\l}a, and Z.~Szyma\'{n}ski, Phys.\ Rev.\ C {\bf 63}, 024308 (2001).

\bibitem{Syed09} N. U. H. Syed, M. Guttormsen, F. Ingebretsen, A. C. Larsen, T. L\"onnroth, J. Rekstad, A. Schiller, S. Siem, and A. Voinov, Phys. Rev. C {\bf 79}, 024316 (2009).


\bibitem{ToI} R. Firestone and V. S. Shirley, {\em Table of Isotopes}, 8th ed. (Wiley, New York, 1996), Vol. II.

\bibitem{Fe+Mo_lev} A.~Schiller, E.~Algin, L.~A.~Bernstein, P.~E.~Garrett, M.~Guttormsen, M.~Hjorth-Jensen, C.~W.~Johnson, G.~E.~Mitchell, J.~Rekstad, S.~Siem, A.~Voinov, and W.~Younes, Phys.\ Rev.\ C \bf 68\rm, 054326 (2003).

\bibitem{gutt4} M.~Guttormsen, M.~Hjorth-Jensen, E.~Melby, J.~Rekstad, A.~Schiller, and S.~Siem, Phys.\ Rev.\ C \bf 63\rm, 044301 (2001).

\bibitem{CecSc} A. C. Larsen, M. Guttormsen, R. Chankova, F. Ingebretsen, T.~L\"onnroth, S. Messelt, J. Rekstad, A. Schiller, S. Siem, N.~U.~H.Syed, and A. Voinov, Phys.\ Rev.\ C \bf 76\rm, 044303 (2007).

\bibitem{SyedTi} N. U. H. Syed, A. C. Larsen, A. B\"urger, M. Guttormsen, S.~Harissopulos, M. Kmiecik, T. Konstantinopoulos, M. Krti\v{c}ka, A. Lagoyannis, T. L\"onnroth, K. Mazurek, M. Norby, H.~T.~Nyhus, G. Perdikakis, S. Siem, and A. Spyrou, Phys.\ Rev.\ C \bf 80\rm, 044309 (2009).

\bibitem{MagneProceedings} M.~Guttormsen, U.~Agvaanluvsan, E.~Algin, A.~B\"{u}rger, A.~C.~Larsen, G.~E.~Mitchell, H.~T.~Nyhus, S.~Siem, H.~K.~Toft, and A.~Voinov, \textit{Properties of warm nuclei in the quasi-continuum},
in \textit{Proceedings of the CNR* 09 Conference}, to be published in EPJ Web of Conferences.

\bibitem{RIPL-2} T. Belgya, O. Bersillon, R. Capote, T. Fukahori, G. Zhigang, S. Goriely, M. Herman, A. V. Ignatyuk, S. Kailas, A. Koning, P. Oblozinsky, V. Plujko and P. Young. {\em Handbook for calculations of nuclear reaction data, RIPL-2}. {\bf IAEA-TECDOC-1506} (IAEA, Vienna, 2006). Available online at http://www-nds.iaea.org/RIPL-2/

\bibitem{KappaMy} J. Y. Zhang, N. Xu, D. B. Fossan, Y. Liang, R. Ma, and E. S. Paul, Phys.\ Rev.\ C \bf 39\rm, 714 (1989).

\bibitem{BCS} J. Bardeen, L. N. Cooper, and J. R. Schrieffer, Phys.\ Rev. \bf 108\rm, 1175 (1957).

\bibitem{Krane} K. S. Krane, {\em Introductory Nuclear Physics} (John Wiley \& Sons, 1988).

\bibitem{Alhassid} Y. Alhassid, S. Liu, and H. Nakada, Phys. Rev. Lett. {\bf 99}, 162504 (2007).

\bibitem{Kopecky} J. Kopecky and M. Uhl, Phys.\ Rev.\ C \bf 41\rm, 1941 (1990).

\bibitem{Voinov01} A. Voinov, M. Guttormsen, E. Melby, J. Rekstad, A. Schiller, and S. Siem, Phys.\ Rev.\ C \bf 63\rm, 044313 (2001).

\bibitem{Varlamov09} V. V. Varlamov, B. S. Ishkhanov, V. N. Orlin, V. A. Tchetvertkova, Moscow State Univ. Inst. of Nucl. Phys. Reports No.2009, p.3/847 (2009).

\bibitem{Fultz69} S. C. Fultz, B. L. Berman, J. T. Coldwell, R. L. Bramblett, and M. A. Kelly, Phys.\ Rev. {\bf 186}, 1255 (1969).

\bibitem{Varlamov03} V. V. Varlamov, N. N. Peskov, D. S. Rudenko, and M. E. Stepanov, Vop. At. Nauki i Tekhn., Ser. Yadernye Konstanty {\bf 1-2} (2003).

\bibitem{Lepretre74} A. Lepr\^{e}tre, H. Beil, R. Bergere, P. Carlos, A. De Miniac, A. Veyssiere, and K. Kernbach, Nucl.\ Phys.\ \textbf{A219}, 39 (1974).

\bibitem{Kopecky87} J. Kopecky and R. E. Chrien, Nucl.\ Phys.\ \textbf{A468}, 285 (1987).

\bibitem{KMF} S. G. Kadmenski\u{\i}, V. P. Markushev, and V. I. Furman, and Yad. Fiz. {\bf 37}, 277 (1983) [Sov. J. Nucl Phys. {\bf 37}, 165 (1983)].

\bibitem{Winhold} E. J. Winhold, E. M. Bowey, D. B. Gayther, and B. H. Patrick, Physics Letters {\bf 32B}, 7 (1970).

\bibitem{Goriely98} S. Goriely, Phys. Lett. B {\bf 436}, 10 (1998).

\bibitem{Govaert} K. Govaert, F. Bauwens, J. Bryssinck, D. De Frenne, E. Jacobs, W. Mondelaers, L. Govor, V. Y. Ponomarev, Phys.\ Rev.\ C \bf 57\rm, 2229 (1998).

\bibitem{Tonchev} A. Tonchev (private communication).

\bibitem{Adrich} P. Adrich, A. Klimkiewicz, M. Fallot, K. Boretzky, T. Aumann, D. Cortina-Gil, U. Datta Pramanik, Th. W. Elze, H. Emling, H. Geissel, M. Hellstr\"om, K. L. Jones, J. V. Kratz, R. Kulessa, Y. Leifels, C. Nociforo, R. Palit, H. Simon, G. Sur\'{o}wka, K. S\"ummerer, and W. Walu{\'s}, Phys. Rev. Lett. {\bf 95}, 132501 (2005).

\bibitem{Klimkiewicz} Klimkiewicz {\em et al.}, Phys. Rev. C {\bf 76}, 051603(R) (2007).

\bibitem{Isacker} P. Van Isacker, M. A. Nagarajan, and D. D. Warner, Phys.\ Rev.\ C \bf 45\rm, R13 (1992).

\bibitem{Paar} N. Paar, P. Ring, T. Niksic, D. Vretenar,  Phys. Rev. C {\bf 67}, 034312 (2003).

\bibitem{Sarchi} D. Sarchi {\em et al.}, Phys. Lett. B {\bf 601}, 27 (2004).

\bibitem{Djalali} C. Djalali, N. Marty, M. Morlet, A. Willis, J. C. Jourdain, N. Anantaraman, G. M. Crawley and, A. Galonsky, and P. Kitching, Nucl.\ Phys.\ \textbf{A388}, 1 (1982).


\end{references}
\end{document}